\documentclass[letter,bibyear]{aa} 

\usepackage{epsfig}
\usepackage{amsmath}
\usepackage{subfigure}
\usepackage{natbib}
\usepackage{multirow}
\usepackage{color}
\usepackage{longtable}
\usepackage{graphicx}
\usepackage{epstopdf}
\usepackage{booktabs}
\usepackage{txfonts}

\newcommand{\jmst}{J.~Mol.~Struct.}   

\begin{document}

\title{Discovery of HC$_4$NC in TMC-1: A study of the isomers of HC$_3$N, HC$_5$N, and HC$_7$N 
\thanks{Based on observations carried out with the Yebes 40m telescope
  (projects 19A003, 19A010, 20A014).
  The 40m radiotelescope at Yebes Observatory is operated by the Spanish Geographic Institute
  (IGN, Ministerio de Transportes, Movilidad y Agenda Urbana).}}

\author{
J.~Cernicharo\inst{1},
N.~Marcelino\inst{1},
M.~Ag\'undez\inst{1},
C.~Berm\'udez\inst{1}
C.~Cabezas\inst{1},
B.~Tercero\inst{2,3},
and J.~R.~Pardo\inst{1}
}

\institute{Grupo de Astrof\'isica Molecular, Instituto de F\'isica Fundamental (IFF-CSIC), 
C/ Serrano 121, 28006 Madrid, Spain. \email jose.cernicharo@csic.es
\and Centro de Desarrollos Tecnol\'ogicos, Observatorio de Yebes (IGN), 19141 Yebes, Guadalajara, Spain.
\and Observatorio Astron\'omico Nacional (OAN, IGN), Madrid, Spain.
}

\date{Received;Accepted}

\abstract{We present a study of the isocyano isomers of the cyanopolyynes
  HC$_3$N, HC$_5$N, and HC$_7$N in TMC-1 and IRC+10216 carried out with the
  Yebes 40m radio telescope. This study has enabled us to report the detection, for the first time in space,
  of HCCCCNC in TMC-1 and to give upper limits for HC$_6$NC in the same source. In addition, the
  deuterated isotopologues of HCCNC and HNCCC were detected, along with all
  $^{13}$C substitutions of HCCNC, also for the first time in space. 
  The abundance ratios of HC$_3$N and HC$_5$N, with their isomers, are
  very different in TMC-1 and IRC+10216, namely, {\textit N}(HC$_5$N)/{\textit N}(HC$_4$NC) is $\sim$300 
  and $\ge$2100, respectively. 
  We discuss the chemistry of the metastable isomers of cyanopolyynes in terms of the most likely formation pathways and by comparing observational 
  abundance ratios between different sources.
}


\titlerunning{HCCCCNC in TMC-1}
\authorrunning{Cernicharo et al.}

\maketitle

\section{Introduction}

Millimeter-wave observations of lines from isomers of abundant molecular species provide key
information for studying the chemical paths leading to their formation in interstellar and
circumstellar clouds, an important piece of the astrochemical puzzle. The best known example is
that of HCN and HNC, which are formed in interstellar clouds by the dissociative recombination
of a common species, the HCNH$^+$ cation, in a branching ratio of 1:1.
Once these species are formed, they undergo various reactions that can lead
to changes in the initial abundances, in particular, reactions of HNC with atoms and radicals
(see, e.g. \citealt{Hacar2020} and references therein).
In circumstellar clouds, HCN and HNC are produced very close to the photosphere of the central star under
thermodynamical chemical equilibrium with a branching ratio of $\sim$1000:1 \citep{Cernicharo2013}. 
While HCN maintains its abundance across the envelope, HNC disappears
very quickly when moving away from the star. It reappears again in the zone where galactic UV photons
are able to penetrate the envelope, initiating a very rich photochemistry \citep{Cernicharo2013}.

Some isomers may  not necessarily be formed from a common precursor but as an effect of radiation 
as it is thought to occur for the $cis$ conformer of HCOOH in the Orion Bar photodissociation region \citep{Cuadrado2016}. 
This is an interesting case, in which the absorption of a UV photon by the $trans$ conformer leads, through 
radiative decay, to the $cis$ conformer. This photochemical switch can only work in regions with enough UV 
illumination. In fact, $cis$ HCOOH has been detected in the cold dark clouds B5 and L483 \citep{Taquet2017,Agundez2019}. 
The lack of UV photons in these environments together with their derived lower $cis$-to-$trans$ abundance
ratios, as compared to the Orion Bar, indicate that a different mechanism must operate in these cold clouds.

However, most of the isomers known in space involve radical CN. In addition to HCN and HNC, 
the isocyanide isomers of CH$_3$CN \citep{Cernicharo1988}, HC$_3$N \citep{Kawaguchi1992a,Kawaguchi1992b},
and NCCN \citep{Agundez2018} have been detected in space. HCCNC and HNCCC, the two isomers of HC$_3$N, 
have also been detected towards the circumstellar envelope IRC+10216 
\citep{Gensheimer1997a,Gensheimer1997a}. A comparative study of the abundances of isomers in
interstellar and circumstellar clouds provide an opportunity to understand the chemical
processes leading to their formation, as the chemistry prevailing in cold interstellar clouds
(ion-neutral reactions) and in external layers of circumstellar envelopes (radical-neutral
and photochemical reactions) are very different. In this letter, we present a systematic
study of the HC$_3$N, HC$_5$N, and HC$_7$N isomers, along with their
isotopologues, towards TMC-1 and IRC+10216 making use of the observations described in
Section \ref{observations}. In Section \ref{results} we report on the first detection in space
of HC$_4$NC\footnote{After the original submission of this manuscript and prior to its 
review, we learned about a parallel effort by Xue et al. (2020) using the GBT at lower frequencies.
Their stacked data of three lines present a convicent detection of HC$_4$NC in TMC-1. The 
column density they derive is consistent with that from our more complete study of the isomers,
isotopologues, and chemistry of cyanopolyynes in this source and toward IRC+10216.}, Isocyanodiacetylene, 
which is the most stable isomer after HC$_5$N. This new
molecular species has been detected only towards TMC-1, whereas a very low upper limit has
been obtained towards IRC+10216, which is a rather interesting finding and which we discuss in Section \ref{discussion}.

\section{Observations}
\label{observations}

New receivers, built
within the Nanocosmos
project\footnote{\texttt{https://nanocosmos.iff.csic.es/}} and installed at the 
Yebes 40m radio telescope, (hereafter, Yebes 40m) were used for the observations
of TMC-1. The Q-band receiver consists of two HEMT cold amplifiers covering the
31.0-50.3 GHz band with horizontal and vertical polarizations. Receiver temperatures
vary from 22 K at 32 GHz to 42 K at 50 GHz. The spectrometers are $2\times8\times2.5$ GHz
FFTs with a spectral resolution of 38.1 kHz providing the whole coverage of the Q-band
in both polarizations. The main beam efficiency varies from 0.6 at 32 GHz to 0.43 at 50 GHz. The
beam size of the telescope is 56$''$ and 36$''$ at 31 and 50 GHz, respectively. 
Pointing errors are always within 2-3$''$.

The observations leading to the  Q-band line survey towards TMC-1 were performed in several sessions,
between November 2019 and February 2020. The observing procedure was
frequency switching with a frequency throw of 10\,MHz. The nominal spectral
resolution of 38.1 kHz was left unchanged for the final spectra. 
{The sensitivity, $\sigma$, along the Q-band varies between $\sim$0.6 (31 GHz), $\sim$1.0 (43 GHz), and 
$\sim$2.5 mK (50 GHz). It was derived by removing a polynomial baseline in velocity windows of -6.2 to 
17.8 km\,s$^{-1}$ centered on each observed line.

The intensity scale, antenna temperature
(T$_A^*$), was calibrated using two absorbers at different temperatures and the
atmospheric transmission model (ATM, \citealt{Cernicharo1985, Pardo2001}).
Calibration uncertainties are estimated to be within 10~\%.
All data were analyzed using the GILDAS package\footnote{\texttt{http://www.iram.fr/IRAMFR/GILDAS}}.
The observations in the Q-band of IRC+10216 were previously described by \citet{Cernicharo2019}
and \citet{Pardo2020}.

\section{Results} 
\label{results}
The sensitivity of our observations towards TMC-1 (see section \ref{observations}) is a factor of 5-10 better than
previously published line surveys of this source at the same frequencies \citep{Kaifu2004}. This has allowed us to detect a
forest of weak lines, most of them arising from the isotopologues
of abundant species such as HC$_3$N and HC$_5$N. For the observed line parameters of HC$_3$N and its
isotopologues, see  Appendix \ref{isotopic_abundances}.
With regard to HC$_5$N, in addition to its deuterated species, it has five $^{13}$C and one $^{15}$N isotopologues and
is responsible for 50 features detected within our survey with a signal to noise
ratio (S/N) between 10 for HC$_5^{15}$N, 20-40 for the $^{13}$C isotopologues, and $>$1000 for the main
isotopologue (see also Appendix \ref{isotopic_abundances} for their observed line parameters).
From the seven rotational transitions of HC$_5$N observed within the 
Q-band ({\textit J$_{up}$}=12 to {\textit J$_{up}$}=18),  
we obtained a local standard of rest velocity of the source, namely, v$_{LSR}$, of 5.83$\pm$0.01 km s$^{-1}$. 
From the $^{13}$C and $^{15}$N isotopologes of HC$_5$N, which provide 43 different transitions to estimate
this same velocity, we get v$_{LSR}$ = 5.84$\pm$0.01 km s$^{-1}$. Hence, we
adopt a value v$_{LSR}$ of 5.83 km s$^{-1}$ for further frequency determinations in TMC-1.
The value given by \citet{Kaifu2004} is 5.85 km s$^{-1}$, which is practically identical 
to our result within the observational uncertainties. 

\begin{figure}[]
\centering
\includegraphics[scale=0.6,angle=0]{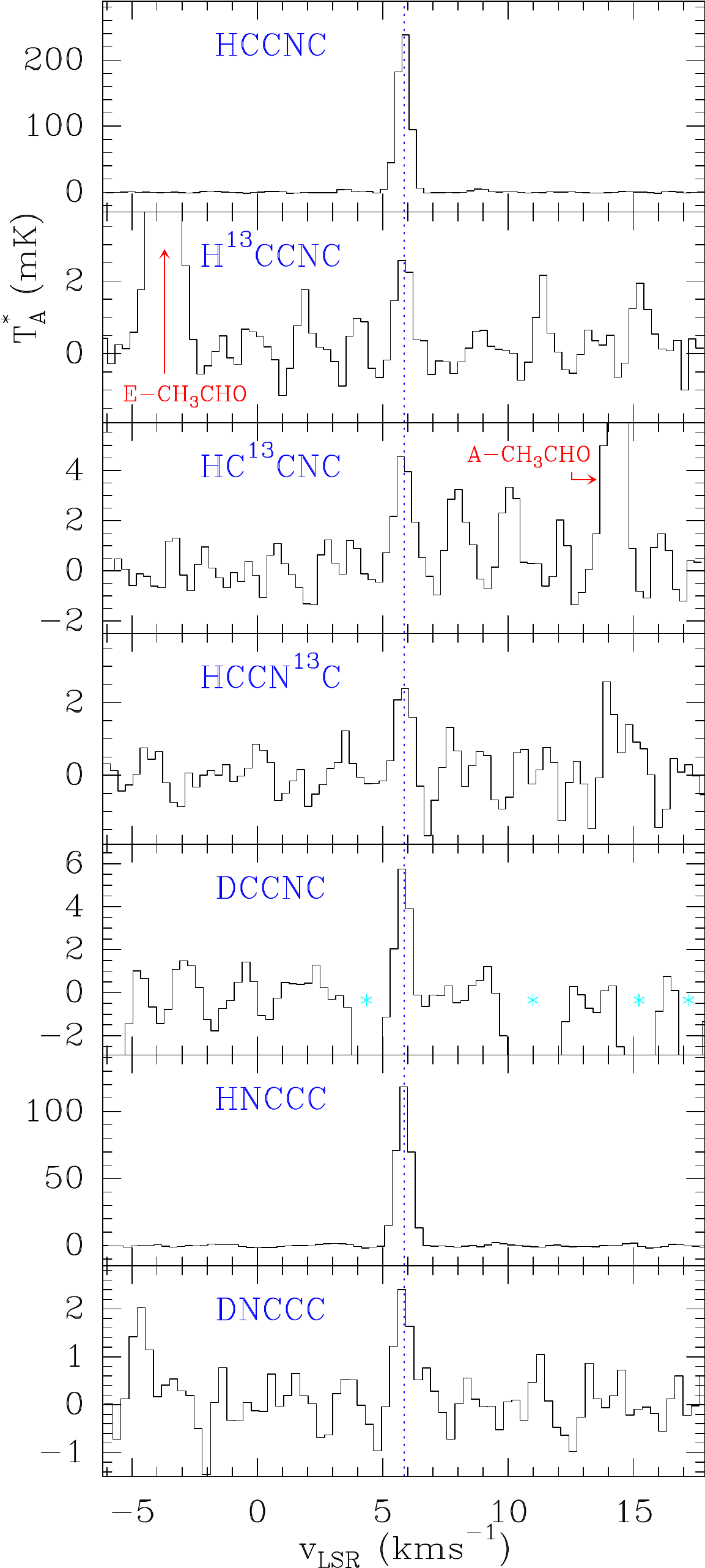}
\caption{$J$=4-3 transition of HNCCC, HCCNC, and some of their
isotopologues, observed towards TMC-1.
The abscissa corresponds to the local standard of rest velocity
in km\,s$^{-1}$. Frequencies and intensities for the observed lines 
are given in Table \ref{tab_hc3n_isomers}.
The ordinate is the antenna temperature corrected for atmospheric and telescope losses in mK.
Cyan stars indicate the position of ghost negative features in the spectra produced by the frequency
switching observation mode. Other spectral features arising from other molecular species, or otherwise unidentified, 
are labeled in the different panels.}
\label{fig_hc3n_isomers}
\end{figure}

\subsection{The isomers of HC$_3$N}
\label{isomers_hc3n}
The most stable isomers of HC$_3$N are HCCNC and HNCCC. They have been extensively observed in the laboratory.
Rotational transitions of HCCNC have been measured up to $J_{up}$=33 at a frequency of 327.8 GHz 
\citep{Guarnieri1992}. HNCCC is a 
quasi-linear species, with the hydrogen atom slightly bent with respect to the NCCC axis. Its
rotational transitions have been measured
in the microwave domain \citep{Hirahara1993} and at millimeter wavelengths \citep{Vastel2018}. Hence,
accurate frequencies are available for these species. The $J$=1-0 line of nine HCCNC isotopologue (D, $^{13}$C
and $^{15}$N) was measured in the laboratory by \citet{Kruger1992}. The millimeterwave
spectrum of DCCNC was measured by \citet{Huckauf1998} up to $J_{up}$=51. The microwave spectrum
of DNCCC was measured by \citet{Hirahara1993} up to $J_{up}$=2, and $\nu_{max}$=19 GHz.
The derived rotational constants for all these species have been implemented in the MADEX code 
\citep{Cernicharo2012} in order to work for their detection and analysis in interstellar and
circumstellar sources.

The HCCNC and HNCCC isomers were previously detected towards TMC-1 
by \citet{Kawaguchi1992a,Kawaguchi1992b} and in IRC\,+10216 by \citet{Gensheimer1997a,Gensheimer1997b}. 
A detailed study of these two species in L1544, including their relative abundances, formation paths, and comparison with the
abundances observed in TMC-1 has been conducted by 
\citet{Vastel2018} based on observations in the $\lambda$ 3 mm domain. 

The derived line parameters for the HCCNC and HNCCC transitions observed in TMC-1 
are given in Table \ref{tab_hc3n_isomers}. Selected lines are shown in Fig. 
\ref{fig_hc3n_isomers}. The deuterated counterparts DNCCC and DCCNC, together with the three $^{13}$C
substitutions of HCCNC, have also been detected for the first time in space. Its derived 
line parameters are given in Table \ref{tab_hc3n_isomers} (see also Fig. \ref{fig_hc3n_isomers}). 
Improved rotational constants for DNCCC, H$^{13}$CCNC, HC$^{13}$CNC, and HCCN$^{13}$C were obtained
from the observed frequencies. They are given in Appendix \ref{App_rot_const}. The isotopic abundance
ratios of the HC$_3$N isomers are discussed in Appendix \ref{isotopic_abundances}.

Assuming a TMC-1 source size of 40$''$ and that the rotational temperature of the observed 
transitions is identical to the kinetic temperature (10 K), we derive
{\textit N}(HNCCC)=(5.2$\pm$0.3)$\times$10$^{11}$ cm$^{-2}$, and 
{\textit N}(HCCNC)=(3.0$\pm$0.3)$\times$10$^{12}$ cm$^{-2}$. We can check the validity of the
assumed rotational temperature by adopting for the isomers the same collisional rates of  
HC$_3$N \citep{Wernli2007}, and a value of (4-10)$\times$10$^4$ cm$^{-3}$ for the H$_2$ volume
density, in the large velocity gradient (LVG) module  
integrated in MADEX \citep{Cernicharo2012}. For the lowest value of the H$_2$ density, the derived
excitation temperatures for the $J$=4-3 and 5-4 transitions are 9.5 and 8.2 K for HCCNC, whereas they are
10 and 7 K for HNCCC. For {\textit n}(H$_2$)=10$^5$ cm$^{-3}$, these rotational transitions will
have excitation temperatures very close to 10 K. For HC$_3$N, we also performed LVG calculations
leading to derived excitation temperatures very close to 10 K for the $J$=4-3 and 5-4 transitions. However,
these HC$_3$N transitions exhibit line opacity problems due to the high abundance of this molecule in TMC-1.
The very weak hyperfine 
components $F$=4-4 and $F$=3-3 of the $J$=4-3 transition, and the $F$=5-5 and $F$=4-4 of the $J$=5-4 transition,  
were detected. We used those components to derive
{\textit N}(HC$_3$N)=(2.3$\pm$0.2)$\times$10$^{14}$ cm$^{-2}$, which is a 
value 1.5 times larger than the one obtained using only the strongest
hyperfine components of the observed rotational transitions. 
We can also estimate the column density of cyanoacetylene by using its $^{13}$C isotopologues
(see Table \ref{tab_hc3n_TMC-1})
and a $^{12}$C/$^{13}$C abundance ratio of 93$\pm$10 (derived from all
isotopologues of HC$_5$N, see Appendix \ref{isotopic_abundances}). In this case, the result is 
{\textit N}(HC$_3$N)=(1.9$\pm$0.2)$\times$10$^{14}$ cm$^{-2}$. Our value for the isotopic abundance
ratio agrees rather well with the one derived by \citet{Takano1998} in the same source.
The derived column densities for HNCCC, HCCNC, and HC$_3$N in TMC-1 (see Table \ref{tab_col_densities})
are in reasonable agreement with those obtained previously \citep{Kawaguchi1992a,Kawaguchi1992b,Takano1998}. 
We note  that our assumed rotational temperatures are higher than those reported in the
literature (5-7 K), but the differences in the estimated column densities for optically thin lines
involving energy levels between 5-8 K is is rather small. For example, for T$_{rot}$=5 K and
N(HNCCC)=10$^{11}$ cm$^{-2}$, the expected brightness intensity for its $J$=4-3 transition is
$\sim$43.5 mK, while for T$_{rot}$=10 K, it is 50.4 mK.

\begin{table}
\small
\caption{Column densities and abundance ratios for the isomers of HC$_3$N and HC$_5$N
towards TMC-1 and IRC+10216.}
\label{tab_col_densities}
\centering
\begin{tabular}{{lcc}}
\hline 
Molecule                                  &     TMC-1           &    IRC+10216 \\
\hline                                                      
{\textit N}(HC$_3$N)                      &(2.3$\pm$0.2)$\times$10$^{14}$&(4.5$\pm$0.2)$\times$10$^{14}$                     \\
{\textit N}(HCCNC)                        &(3.0$\pm$0.3)$\times$10$^{12}$&(1.1$\pm$0.1)$\times$10$^{12}$                     \\
{\textit N}(HNCCC)                        &(5.2$\pm$0.3)$\times$10$^{11}$&(3.4$\pm$0.2)$\times$10$^{11}$                     \\
{\textit N}(HC$_5$N)                      &(1.8$\pm$0.2)$\times$10$^{14}$&(4.2$\pm$0.4)$\times$10$^{14}$ $^b$ \\
{\textit N}(HC$_4$NC)                     &(3.0$\pm$0.7)$\times$10$^{11}$&  $\le$2.1$\times$10$^{11}$   \\
{\textit N}(HNC$_5$)                      &$\le$1.5$\times$10$^{11}$     &  $\le$1.1$\times$10$^{11}$   \\
{\textit N}(HC$_7$N)                      &(6.4$\pm$0.2)$\times$10$^{13}$&(1.9$\pm$0.2)$\times$10$^{14}$ $^c$ \\
{\textit N}(HC$_6$NC)                     & $\le$9.0$\times$10$^{10}$    &  $\le$2.0$\times$10$^{11}$ \\
\hline                                                                            
{\textit N}(HC$_3$N)/{\textit N}(HCCNC)   &  77.0$\pm$8.0      &    392$\pm$22   \\
{\textit N}(HC$_3$N)/{\textit N}(HNCCC)   &  442.0$\pm$70.0    &    1305$\pm$45  \\
{\textit N}(HCCNC)/{\textit N}(HNCCC)     &  5.8$\pm$1.0       &    4.0$\pm$0.9      \\
{\textit N}(HC$_5$N)/{\textit N}(HC$_4$NC)&  600$\pm$70.0      &    $\ge$2000        \\
{\textit N}(HC$_5$N)/{\textit N}(HNC$_5$) &  $\ge$1200         &    $\ge$3800        \\
{\textit N}(HC$_4$NC)/{\textit N}(HNC$_5$)&  $\ge$2            &    $\ge$1.9         \\
{\textit N}(HC$_7$N)/{\textit N}(HC$_6$NC)&  $\ge$710.0        &    $\ge$950         \\
{\textit N}(HC$_3$N)/{\textit N}(HC$_5$N) &  1.3$\pm$0.2       &    1.1$\pm$0.2      \\
{\textit N}(HC$_5$N)/{\textit N}(HC$_7$N) &  2.8$\pm$0.3       &    2.2$\pm$0.5\\

\hline 
\end{tabular}
\tablefoot{\\
        \tablefoottext{a}{All column densities are given in units of cm$^{-2}$.}
        \tablefoottext{b}{For IRC+10216, we adopted
     the value derived by \citet{Pardo2020}, which corresponds to the
     gas component with {\textit T$_{rot}$}=10.1 K. A second component, with similar
     column density, was found for high-$J$ levels with a {\textit T$_{rot}$}=24.5 K.}
        \tablefoottext{c}{For IRC+10216, we adopted
     the value derived by \citet{Pardo2020}, which corresponds to the
     gas component with {\textit T$_{rot}$}=16.8 K. A second component, with a
     column density 2.5 lower, was found for high-$J$ levels with a {\textit T$_{rot}$}=31.6 K.}
     }
\end{table}
\normalsize

The Q-band survey of IRC+10216 carried out with the Yebes 40m also reached an unprecedented sensitivity, which has allowed the detection
of new molecular species such as MgC$_3$N and MgC$_4$H \citep{Cernicharo2019}, and of previously undetected
vibrational excited states of abundant species such as HC$_7$N and HC$_9$N \citep{Pardo2020}. In these
data, we find two transitions of HNCCC and HCCNC, with their observed line parameters, which are given together with
those of HC$_3$N and their isotopologues in Table \ref{tab_hc3n_isomers_irc}. The isotopologues
of the isomers are, however, below the sensitivity limit of the data. We
performed an LVG model to derive the column densities of the three HC$_3$N isomers from
the observations. The results are given in Table \ref{tab_col_densities}. To avoid line
opacity effects, the column density of HC$_3$N has been derived from the observed intensities 
of its three $^{13}$C isotopologues using an isotopic abundance ratio of 45 \citep{Cernicharo2000}.

\subsection{The isomers of HC$_5$N. Detection of HC$_4$NC}
\label{isomers_hc5n}

HC$_5$N has different isomers shown in Figure \ref{ener_isomers}. The most stable one, after
HC$_5$N, is HC$_4$NC (see Appendix \ref{App_ab_initio}). Both of them could be formed in TMC-1
through the dissociative recombination of the cation HC$_5$NH$^+$, recently detected
in TMC-1 by \citet{Marcelino2020}. Through its similarity to HC$_3$N and HNCCC, the isomer HNC$_5$ could
also be formed in the dissociative recombination of HC$_5$NH$^+$. While HC$_4$NC has
been observed in the laboratory \citep{Botschwina1998} and precise rotational frequencies are available 
from the Cologne Database for Molecular Spectroscopy (CDMS) catalogue \citep{Muller2005} or from the MADEX code \citep{Cernicharo2012}, no laboratory data
are available for HNC$_5$. Nevertheless, we  estimated, using ab initio calculations,
its rotational constants and searched for this species in TMC1 and IRC+10216.

\begin{figure}[]
\centering
\includegraphics[scale=0.65,angle=0]{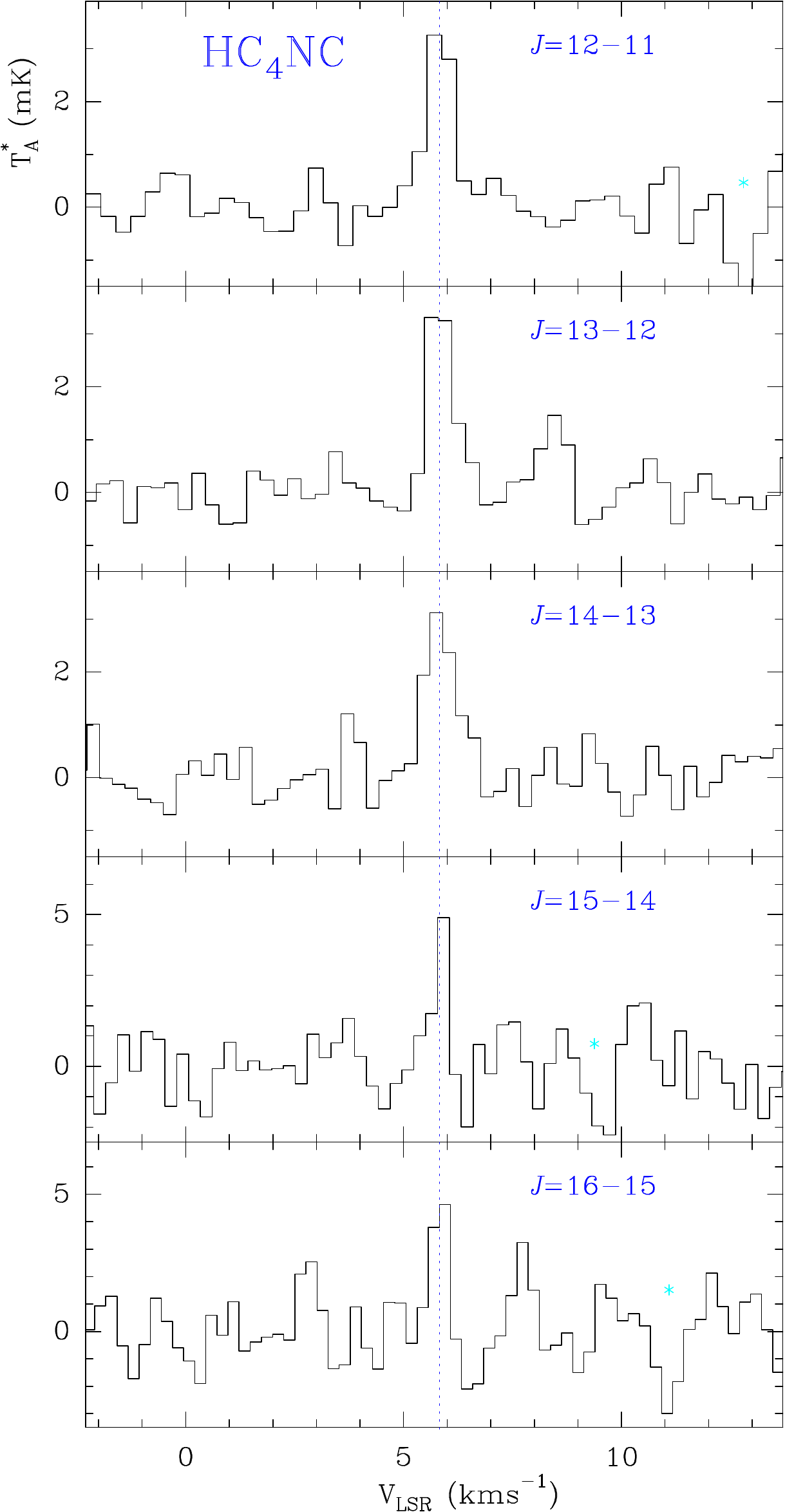}
\caption{Observed lines of HC$_4$NC found in the 31-50 GHz frequency range towards TMC-1.
The abscissa corresponds to the local standard of rest velocity
in km s$^{-1}$. Frequencies and intensities for the observed lines are given in Table \ref{tab_hc4nc}.
The ordinate is the antenna temperature corrected for atmospheric and telescope losses in mK. The {\textit J=17-16}
line is not detected within a 3$\sigma$ level of 5.7 mK. Spectral resolution is 38.1 kHz.}
\label{fig_hc4nc}
\end{figure}

\begin{table}
\caption{Observed line parameters of HCCCCNC in TMC-1.}
\label{tab_hc4nc}
\centering
\begin{tabular}{{ccccc}}
\hline \hline
{\textit J$_u$}& $\nu_{obs}^a$ & T$_A^*$$^b$ & $\Delta$v$^c$ &$\int$T$_A^*$dv $^d$ \\
               &  (MHz)        & (mK)        & (kms$^{-1}$)  &(mK kms$^{-1}$)      \\
\hline
12 & 33628.151$\pm$0.010 &  3.6$\pm$0.5& 0.71$\pm$0.12 & 2.68$\pm$0.3 \\
13 & 36430.429$\pm$0.010 &  3.8$\pm$0.5& 0.66$\pm$0.09 & 2.68$\pm$0.3 \\
14 & 39232.730$\pm$0.010 &  3.1$\pm$0.6& 0.85$\pm$0.12 & 2.76$\pm$0.3 \\
15 & 42035.040$\pm$0.010 &  3.7$\pm$1.0& 0.47$\pm$0.14 & 2.08$\pm$0.4 \\
16 & 44837.295$\pm$0.010 &  4.9$\pm$1.2& 0.42$\pm$0.19 & 2.24$\pm$0.5 \\
\hline
\end{tabular}
\tablefoot{\\
        \tablefoottext{a}{Observed frequencies for a v$_{LSR}$ of 5.83 km s$^{-1}$. The $J$=17-16 line
     has not been detected within a 3$\sigma$ level of 5.7 mK.}
        \tablefoottext{b}{Antenna temperature in mK.}
        \tablefoottext{c}{Linewidth at half intensity derived by fitting a Gaussian line profile to the observed
     transitions (in kms$^{-1}$).}
        \tablefoottext{d}{Integrated line intensity in mK kms$^{-1}$. }
}
\end{table}

Five lines of HC$_4$NC were detected towards TMC-1 (see Fig \ref{fig_hc4nc}). This is the first time that this species has been 
detected in space. From the rotational diagram obtained with its observed line intensities, we derived
{\textit T$_{rot}$}=9.8$\pm$0.9 K, and {\textit N}(HC$_4$NC)=(3.0$\pm$0.7)$\times$10$^{11}$ cm$^{-2}$. In order to
compare the derived column density with that of the most stable isomer, HC$_5$N, we used the observed
intensities of this last species (see Table \ref{tab_hc5n_TMC-1}) to build another rotational diagram. We obtained
{\textit T$_{rot}$}=8.6$\pm$0.2 K, and {\textit N}(HC$_5$N)=(1.8$\pm$0.2)$\times$10$^{14}$ cm$^{-2}$ (corrected for
line opacity effects, see Appendix \ref{isotopic_abundances}). 
Hence, the HC$_5$N/HC$_4$NC abundance ratio in TMC-1 is $\sim$600. Using the observed frequencies of HC$_4$NC in TMC-1 and those obtained in the
laboratory \citep{Botschwina1998}, we improved
the rotational constants of this species (see Appendix \ref{App_rot_const}).

We searched
for the lines of HC$_4$NC towards IRC+10216 without success. By averaging all the data from the expected
line positions within the Q-band, we derived a column density upper limit 
of 2$\times$10$^{11}$ cm$^{-2}$. Using the same dataset, the column density of HC$_5$N was recently 
derived as (4.2$\pm$0.4)$\times$10$^{14}$ cm$^{-2}$ \citep{Pardo2020}.
Hence, in this carbon-rich circumstellar envelope, the HC$_5$N/HC$_4$NC abundance ratio is $\ge$2000. 

Precise rotational constants were estimated through our ab initio calculations
for HNC$_5$ (scaled values $B_0$=1358.8$\pm$1.0 MHz, $D_0\sim$32 Hz; see Appendix \ref{App_ab_initio}). 
Although slightly asymmetrical, it is a quasi-linear species
(see Appendix \ref{App_ab_initio}) such that we could expect the
$J_{0J}$ series of lines to be in harmonic relation. The $K$=1 lines could be at an energy too
high to be detected in TMC1. We searched for such an harmonic pattern around 
$\pm$100 MHz of the predicted
line frequencies without success. Using the dipole moment we derived from our
calculations, we get {\textit N}(HNC$_5$)$\le$1.5$\times$10$^{11}$ cm$^{-2}$ and  $\le$1.1$\times$10$^{11}$ cm$^{-2}$ towards TMC-1 
and IRC+10216, respectively.

\subsection{The isomers of HC$_7$N}
HC$_7$N has 17 rotational transitions within the frequency range of our Q-band data.
All of them have been detected in TMC1 and in IRC+10216. For the latter source, the data
were presented and analyzed by \citet{Pardo2020}; the derived HC$_7$N column density is
given in Table \ref{tab_col_densities}. From the TMC1 data (see Table \ref{tab_hc5n_TMC-1}),
we built a rotational diagram that gives the following results: {\textit T$_{rot}$}=7.6$\pm$0.1 K 
and {\textit N}(HC$_7$N)=(6.4$\pm$0.4)$\times$10$^{13}$ cm$^{-2}$.

The most stable isomer of HC$_7$N is HC$_6$NC, which was previously observed in the laboratory
by \citet{Botschwina1998}. We searched for its lines towards both sources without success.
The derived upper limits, which are based on a stacking of the $J$=28-27 up to $J$=31-30 transitions, 
are given in Table \ref{tab_col_densities}.

\section{Discussion}
\label{discussion}

The chemistry of HC$_3$N isomers in cold dense clouds has been discussed by 
\cite{Osamura1999} and \cite{Vastel2018}. In the scenario depicted by these 
authors, the main formation route to the HC$_3$N isomers relies on the dissociative 
recombination of HC$_3$NH$^+$. The rate constant of this reaction has been 
measured, using the deuterated variant of the ion, by \cite{Geppert2004}. Calculations 
by \cite{Osamura1999} indicate that rearrangement is possible so that the various linear 
or nearly-linear isomers of HC$_3$N, that is, HNC$_3$, HCCNC, and HCNCC, can be formed. There 
is some experimental information on the different fragments that are formed \citep{Geppert2004}, but 
precise branching ratios for the different exit channels are not known and introduce important 
uncertainties when modeling the chemistry of HC$_3$N isomers. The chemical models constructed 
for TMC-1 \citep{Osamura1999} and for L1544 \citep{Vastel2018} reproduce quite well the 
observed abundances of HC$_3$N isomers, although they tend to overestimate the abundances 
of HNC$_3$ and HCCNC with respect to HC$_3$N. In these models, the most stable isomer, HC$_3$N, 
is also formed through neutral-neutral reactions, while the different metastable isomers are 
mostly formed through dissociative recombination of HC$_3$NH$^+$ and, to a lesser extent, of 
its less stable isomer HCCNCH$^+$. The higher abundance observed for the isomer HCCNC with 
respect to HNC$_3$ is explained in terms of a higher reactivity of the latter isomer with 
neutral H and C atoms \citep{Osamura1999}. 
By way of an analogy with HC$_3$N, the main source of HC$_5$N and its isomers is most likely the 
dissociative recombination of HC$_5$NH$^+$, recently detected in TMC-1 \citep{Marcelino2020}. 
However, branching ratios for the formation of the different isomers HC$_5$N, HC$_4$NC, 
and HNC$_5$ are unknown.

\begin{figure}[]
\centering
\includegraphics[scale=0.55,angle=0]{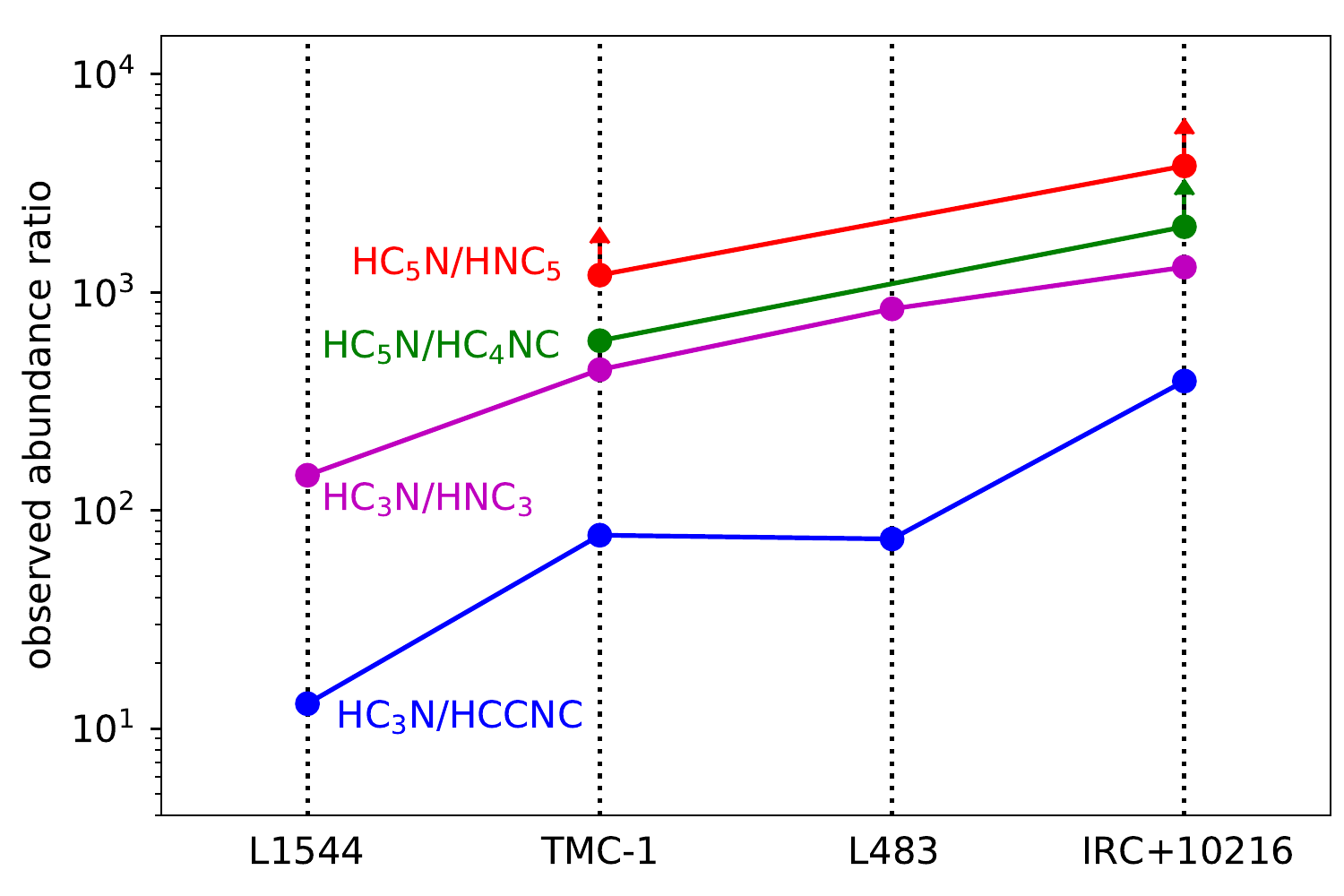}

\caption{Observed abundance ratios between isomers of HC$_3$N and HC$_5$N in different molecular 
sources. Values for TMC-1 and IRC\,+10216 are from this study, for L1544 from \cite{Vastel2018}, and 
for L483 from \cite{Agundez2019}.}
\label{fig_ratios}
\end{figure}

Looking at how the relative abundances of HC$_3$N and HC$_5$N isomers behave in different sources can 
provide clues on the underlying chemical processes. In Fig.~\ref{fig_ratios}, we compare the abundance 
ratios between HC$_3$N isomers derived here for TMC-1 and IRC\,+10216 with those obtained in the molecular 
clouds L1544 and L483. It can be seen that the HC$_3$N/HCCNC and HC$_3$N/HNC$_3$ ratios observed in TMC-1 are 
similar to those seen in the Class\,0 molecular cloud L483, although they are higher than those derived in 
the starless core L1544 and lower than the values in the C-star envelope IRC\,+10216. The change in these 
abundance ratios may be related to the temperature of the source. The low temperature of the starless core 
L1544 seems to favor the formation of metastable isomers while these are less efficiently formed in the 
warmer ejecta of IRC\,+10216. Indeed, as temperature increases, the presence of high-energy isomers is 
less favorable because chemical reactions, including isomerization, are expected to favor the most 
stable isomer. The same is expected to be true of isotopic fractionation. 
For example, in TMC-1, the 
isotopologue HCC$^{13}$CN is twice as abundant as H$^{13}$CCCN and HC$^{13}$CCN (see 
also \citealt{Takano1998}), whereas, in the case of HCCNC, its HC$^{13}$CNC isotopologue is 1.8 times more abundant 
than H$^{13}$CCNC and HCCN$^{13}$C (see Table \ref{col_den_iso_isomers}). A slight overabundance of HCCCC$^{13}$CN with respect to the other four $^{13}$C isotopologues of HC$_5$N 
has also been observed (see Appendix \ref{isotopic_abundances}). These kinds of isotopic anomalies have been also 
seen for HC$_3$N in other cold dense clouds \citep{Araki2016,Taniguchi2017,Agundez2019}, but they are not seen 
in IRC\,+10216.

The column densities and abundance ratios derived for the isomers of HC$_3$N, HC$_5$N, and HC$_7$N in TMC-1 
and IRC\,+10216 are summarized in Table \ref{tab_col_densities}. The abundance ratios given in Table 
\ref{tab_col_densities} (see also Fig.~\ref{fig_ratios}), suggest a trend in which the relative abundance 
of metastable isomers decreases as we move to larger molecular sizes. In TMC-1, we have $N$(HCN)/$N$(HNC)$\simeq$1, 
$N$(HC$_3$N)/$N$(HCCNC)$\simeq$77, $N$(HC$_3$N)/$N$(HNC$_3$)$\simeq$442, $N$(HC$_5$N)/$N$(HC$_4$NC)$\sim$600, and 
$N$(HC$_7$N)/$N$(HC$_6$NC)$\ge$710. These ratios are increased by a factor of $\sim$4-5 in IRC\,+10216. This suggests 
that metastable isomers of longer carbon chains will be hardly detectable in these sources.

\begin{acknowledgements}

The Spanish authors thank Ministerio de Ciencia e Innovaci\'on for funding
support through project AYA2016-75066-C2-1-P. We also thank ERC for funding through grant
ERC-2013-Syg-610256-NANOCOSMOS. MA and CB thanks Ministerio de Ciencia e Innovaci\'on
for Ram\'on y Cajal grant RyC-2014-16277 and Juan de la Cierva grant FJCI-2016-27983.
\end{acknowledgements}

\normalsize

\begin{appendix}
\section{Improved rotational constants for the isotopologues of HNCCC and HCCNC and for HC$_4$NC}
\label{App_rot_const}
While the rotational spectrum of HCCNC was already measured in the laboratory up to
$J_{up}$=33 by \citet{Guarnieri1992}, only the $J$=1-0 rotational transition 
of its $^{13}$C isotopologues has been measured so far \citep{Kruger1992}.
Predictions for the frequencies of their rotational transitions have been obtained 
by initially assuming a distortion constant
similar to that of the main isotopologue. This allows us to derive frequencies with
uncertainties of 50-140 kHz for lines in the 31-50 GHz frequency interval. The $J$=4-3 and 5-4 lines
of H$^{13}$CCNC, HC$^{13}$CNC, and HCCN$^{13}$C were easily detected in TMC-1 and their
frequencies have been obtained with an accuracy better than 10 kHz assuming
a v$_{LSR}$ for the source of 5.83 km\,s$^{-1}$ (see Section \ref{results}). The derived
line parameters are given in Table \ref{tab_hc3n_isomers} and the spectra are
shown in Fig. \ref{fig_hc3n_isomers}. The observed line parameters for these species
in IRC+10216 are given in Table \ref{tab_hc3n_isomers_irc}

\begin{table*}
\caption{Observed line parameters of HNCCC and HCCNC in TMC-1.}
\label{tab_hc3n_isomers}
\centering
\begin{tabular}{{|l|c|c|c|c|c|c|}}
\hline \hline
Molecule     & {\textit J$_u$-J$_l$}& $\nu_{obs}^a$ & T$_A^*$& $\Delta$v$^b$&$\int$T$_A^*$dv $^c$& $\sigma^d$ \\
             &                      &  (MHz)        & (mK)   & (kms$^{-1}$) &(mK kms$^{-1}$)     & (mK)     \\
\hline
HNCCC        & 4-3 & 37346.541(10) & 118.7          &0.70(1) & 88.8(10)&0.6\\
HNCCC        & 5-4 & 46683.061(10) & 121.2          &0.57(1) & 73.5(10)&1.5\\
DNCCC        & 4-3 & 35204.594(10) &   2.4          &0.76(9) & 1.95(06)&0.5\\
DNCCC        & 5-4 & 44005.628(10) &   3.2          &0.46(9) & 1.64(07)&1.1\\
\hline                                             
\hline                                             
HCCNC        & 4-3 & 39742.549(10) & 249.2          &0.62(1) &164.7(10)&0.8\\
HCCNC        & 5-4 & 49678.075(10) & 224.6          &0.58(1) &138.4(10)&1.9\\
DCCNC        & 4-3 & 36786.182(10) &   6.1$^e$      &0.58(9) & 3.74(90)&0.7\\
DCCNC        & 5-4 & 45982.625(10) &   9.9          &0.60(8) & 6.38(30)&1.1\\
H$^{13}$CCNC & 4-3 & 38504.813(10) &   2.8          &0.68(9) & 2.03(30)&0.6\\
H$^{13}$CCNC & 5-4 & 48130.916(10) &   4.5          &0.45(9) & 2.50(80)&1.7\\
HC$^{13}$CNC & 4-3 & 39595.406(10) &   4.7          &0.80(9) & 3.97(30)&0.7\\
HC$^{13}$CNC & 5-4 & 49494.139(07) &$\le$4.5$^f$    &        &         &1.5\\
HCCN$^{13}$C & 4-3 & 38438.864(10) &   2.7          &0.68(9) & 1.92(40)&0.6\\
HCCN$^{13}$C & 5-4 & 48048.449(10) &   5.4          &0.45(9) & 2.58(70)&1.8\\
HCC$^{15}$NC & 4-3 & 39557.751(50) &$\le$2.2$^g$    &        &         &0.7\\
\hline
\end{tabular}
\tablefoot{\\
        \tablefoottext{a}{Observed frequencies for a v$_{LSR}$ of 5.83 km s$^{-1}$. For all entries, 
     values between parentheses represent the uncertainty in units of the last digit.}\\
        \tablefoottext{b}{Linewidth at half intensity derived by fitting a Gaussian line profile to the observed
     transitions (in kms$^{-1}$).}\\
        \tablefoottext{c}{Integrated line intensity in mK km\,s$^{-1}$. }\\
        \tablefoottext{d}{The sensitivity of the data (root mean square error) has been derived from a baseline to the each line 
      in a velocity window -6.2 to 17.8 km\,s$^{-1}$ (in mK). }\\
        \tablefoottext{e}{Blended with a negative feature produced by the frequency switching observing mode.}\\
        \tablefoottext{f}{3$\sigma$ upper limit. Observed frequency has been replaced by the predicted frequency.}\\
}
\end{table*}

\begin{table}
\caption{Observed line parameters for HNCCC, HCCNC, and HC$_3$N in IRC+10216.}
\label{tab_hc3n_isomers_irc}
\centering
\begin{tabular}{{|l|c|c|c|}}
\hline 
Molecule& {\textit J$_u$-J$_l$}& $\nu_{obs}^a$ & $\int$T$_A^*$dv $^b$ \\
        &                &  (MHz)        & (K kms$^{-1}$)       \\
\hline
HNCCC        &4-3& 37346.448(50)& 0.064(07)\\ 
HNCCC        &5-4& 46683.020(80)& 0.077(07)\\  
\hline                                
\hline                                
HCCNC        &4-3& 39742.549(30)& 0.075(04)\\
HCCNC        &5-4& 49678.155(50)& 0.078(06)\\
\hline                                
\hline                                
HCCCN        &4-3& 36392.327(10)& 21.21(15)\\
HCCCN        &5-4& 45490.311(10)& 24.20(15)\\
H$^{13}$CCCN &4-3& 35267.369(20)& 0.620(08)\\
H$^{13}$CCCN &5-4& 44084.175(50)& 0.959(07)\\
HC$^{13}$CCN &4-3& 36237.909(50)& 0.621(08)\\
HC$^{13}$CCN &5-4& 45297.243(50)& 0.913(07)\\
HCC$^{13}$CN &4-3& 36241.427(50)& 0.605(08)\\
HCC$^{13}$CN &5-4& 45301.689(50)& 1.046(08)\\
\hline
\end{tabular}
\tablefoot{\\
        \tablefoottext{a}{Observed frequencies for a v$_{LSR}$ of -26.5 km s$^{-1}$ \citep{Cernicharo2000}. 
     For all entries, values between parentheses represent the uncertainty in units of the last digit.}\\
        \tablefoottext{b}{Integrated line intensity in K kms$^{-1}$.}\\
}
\end{table}

For the deuterated species, DCCNC, the rotational spectrum up to $J_{up}$=51 was measured by \citet{Huckauf1998}. Its rotational lines within the Q-band
are predicted with a high level of accuracy (the derived line parameters
are given in Table \ref{tab_hc3n_isomers}, see Fig. \ref{fig_hc3n_isomers}). 
For DNCCC, only the $J$=1-0 and 2-1 lines
were measured in the laboratory by \citet{Hirahara1993}. In our study, we detected the $J$=4-3 and 5-4 lines
in TMC-1 and, again, we determined their central frequencies with an accuracy better than 10 kHz.
The derived
line parameters are also given in Table \ref{tab_hc3n_isomers} and the corresponding spectra are
shown in Fig. \ref{fig_hc3n_isomers}. 

As a next step, by combining the laboratory data \citep{Kruger1992,Hirahara1993} with the
frequencies derived towards TMC-1, we obtained a new set of rotational constants
for DNCCC, H$^{13}$CCNC, HC$^{13}$CNC, and HCCN$^{13}$C. The hyperfine structure parameters,
nuclear quadrupole, spin-rotation, and spin-spin constants have been fixed to the
values derived from laboratory data, that is, we fitted only the rotational and
the distortion constants. The results are
provided in Table \ref{new_rot_const}. With this new set of constants, the frequencies of
the isotopologues can be predicted with accuracies better than 20 kHz at rest frequencies
up to 80 GHz, and better than 100 kHz for rest frequencies up to 130 GHz ($J_{up}$=15).

For HC$_4$NC, the laboratory data cover lines from $J$=3-2 up to $J$=6-5 \citep{Botschwina1998}. 
The rotational lines observed in TMC1 add five more rotational transitions, from $J=$12-11 up to $J$=16-15. 
By fitting the laboratory and astrophysical lines, a new set of
rotational constants  were obtained for this species (see Table \ref{new_rot_const}).
The best improvement is obtained for the distortion constant, $D_0$, which is derived to be
32.8$\pm$0.4 Hz, compared with the laboratory value of 34.3$\pm$0.9 Hz. Predictions using the new
constants have an accuracy better than 20 kHz up to 70 GHz, and better than  100 kHz up to 115 GHz
($J$=41-40).

\begin{table}
\caption{New rotational and distortion constants for some isotopologues of HNCCC and HCCNC
and for HC$_4$NC.}
\label{new_rot_const}
\centering
\begin{tabular}{{|l|c|c|c|c|}}
\hline
Molecule       &    $B_0^a$     &  $D_0$    & $J_{max}$& $\sigma^b$\\          
               &    (MHz)     & (kHz)   &          & (kHz) \\
\hline
DNCCC          & 4400.5944(4) & 0.63(9) & 5        & 1.7 \\
H$^{13}$CCNC   & 4813.1187(6) & 0.54(4) & 5        & 2.5 \\
HC$^{13}$CNC   & 4949.4462(3) & 0.64(4) & 4        & 1.0 \\
HCCN$^{13}$C   & 4804.8737(5) & 0.56(3) & 5        &13.8 \\
HCCCCNC$^c$    & 1401.18216(6)& 0.0328(4)& 16      & 7.0 \\
\hline
\end{tabular}
\tablefoot{\\
        \tablefoottext{a}{Values between parentheses represent the uncertainty in units of the last digit.}\\
        \tablefoottext{b}{Standard deviation of the fit.}\\
    \tablefoottext{c}{The fitted value for eQq is 0.97(3) MHz.}\\
}
\end{table}

\section{HC$_3$N, HC$_5$N, and HC$_7$N column densities and isotopic abundance ratios}
\label{isotopic_abundances}
Rest frequencies for the $^{13}$C isotopologues of HC$_3$N
were obtained by fitting all the observed rotational transitions in the
laboratory \citep{Creswell1977,Mallinson1978,Thorwirth2001}. For the $^{15}$N 
isotopologue, the measured frequencies from these references and those from
\citet{Fayt2004} have been used
to derive its rotational parameters. 

For DC$_3$N, all available laboratory data 
were used to derive its rotational parameters
\citep{Mallinson1978,Tack1983,Spahn2008}. The main isotopologue has been extensively
observed in the laboratory up to $J$=89 \citep{deZafra1971,Mbosei2000,Creswell1977,
Chen1991,Yamada1995,Thorwirth2000}. The derived rotational parameters were
implemented in the MADEX code \citep{Cernicharo2012}. The derived rest frequencies
for the observed transitions are given in Table \ref{tab_hc3n_TMC-1}.

For HC$_5$N and its $^{13}$C, $^{15}$N, and D isotopologues
we fit the rotational transitions measured in the laboratory by \citet{Bizzocchi2004}
and, again, the resulting rotational constants have been implemented in the MADEX code. The rest
frequencies for the observed transitions are given in Table \ref{tab_hc5n_TMC-1}.

In order to derive column densities for all isotopologues of HC$_3$N, HNCCC, and HCCNC, 
for which only two rotational lines ($J$=4-3 and 5-4) have been observed in
TMC-1, we assumed that the molecule is thermalized at a kinetic temperature of the cloud of 10 K
and a source radius of 40$''$ (see Section \ref{isomers_hc3n}). 
The derived column densities are given in Table \ref{col_den_iso_isomers} and the
line parameters for the HC$_3$N rotational transitions 
used for this analysis are given in Tables \ref{tab_hc3n_TMC-1}.
We confirm a previous result of \citet{Takano1998} concerning the overabundance of HCC$^{13}$CN with
respect to the other two $^{13}$C isotopologues. These authors analyzed the chemical
processes that could lead to the enriched abundance for this particular isotopologue and
concluded that it is produced during the formation process of HC$_3$N -- and not through a reaction
of $^{13}$C$^+$ with HC$_3$N. This effect also shows up in HCCNC, for which its 
HC$^{13}$CNC isotopologue is 1.8 times more abundant than H$^{13}$CCNC and HCCN$^{13}$C
(see Table \ref{col_den_iso_isomers}). We conclude, hence,
that the carbon-13 attached to the nitrogen atom has a clear overabundance with respect to the
other isotopologues of HCCCN and HCCNC.

\begin{table}
\caption{Derived column densities for the isotopologues of HC$_3$N and HC$_5$N in TMC-1.}
\label{col_den_iso_isomers}
\centering
\begin{tabular}{{|l|c|c|}}
\hline
Species  & {\textit T$_{rot}$$^a$}& {\textit N$^b$}\\
         &         (K)            &   (cm$^{-2}$)   \\
\hline 
HCCCN          & 10.0$^d$ & (2.3$\pm$0.2)$\times$10$^{14}$$^c$\\
H$^{13}$CCCN   & 10.0$^d$ & (2.5$\pm$0.2)$\times$10$^{12}$\\
HC$^{13}$CCN   & 10.0$^d$ & (2.9$\pm$0.2)$\times$10$^{12}$\\
HCC$^{13}$CN   & 10.0$^d$ & (3.6$\pm$0.2)$\times$10$^{12}$\\
DCCCN          & 10.0$^d$ & (3.7$\pm$0.2)$\times$10$^{12}$\\
\hline                            
HCCNC          & 10.0$^d$ & (3.0$\pm$0.2)$\times$10$^{12}$\\
H$^{13}$CCNC   & 10.0$^d$ & (4.1$\pm$0.4)$\times$10$^{10}$\\
HC$^{13}$CNC   & 10.0$^d$ & (7.3$\pm$0.4)$\times$10$^{10}$\\
HCCN$^{13}$C   & 10.0$^d$ & (4.0$\pm$0.4)$\times$10$^{10}$\\  
DCCNC          & 10.0$^d$ & (1.0$\pm$0.3)$\times$10$^{11}$\\   
\hline
HNCCC          & 10.0$^d$ & (5.2$\pm$0.3)$\times$10$^{11}$\\
DNCCC          & 10.0$^d$ & (1.2$\pm$0.2)$\times$10$^{10}$\\
\hline
HC$_5$N        & 8.6$\pm$0.2 & (1.8$\pm$0.2)$\times$10$^{14}$$^c$\\ 
H$^{13}$CCCCCN & 7.3$\pm$0.2 & (1.8$\pm$0.2)$\times$10$^{12}$\\
HC$^{13}$CCCCN & 6.0$\pm$0.1 & (2.1$\pm$0.2)$\times$10$^{12}$\\
HCC$^{13}$CCCN & 6.6$\pm$0.2 & (2.0$\pm$0.4)$\times$10$^{12}$\\
HCCC$^{13}$CCN & 7.8$\pm$0.4 & (1.7$\pm$0.3)$\times$10$^{12}$\\
HCCCC$^{13}$CN & 7.3$\pm$0.2 & (2.1$\pm$0.2)$\times$10$^{12}$\\
HCCCCC$^{15}$N & 8.7$\pm$1.4 & (4.6$\pm$1.0)$\times$10$^{11}$\\
DCCCCCN        & 7.3$\pm$0.4 & (2.2$\pm$0.4)$\times$10$^{12}$\\
\hline
HC$_7$N        & 7.6$\pm$0.1 & (6.4$\pm$0.4)$\times$10$^{13}$\\
\hline
\end{tabular}
\tablefoot{\\
        \tablefoottext{a}{Derived rotational temperature in K from the data in Tables
    \ref{tab_hc3n_TMC-1} and \ref{tab_hc5n_TMC-1}.}\\
        \tablefoottext{b}{Derived column density in cm$^{-2}$.}\\
        \tablefoottext{c}{Corrected for line opacity effects (see text).}\\
        \tablefoottext{d}{Adopted value (see text).}\\
    
}
\end{table}

For IRC+10216, the isotopic abundances have been discussed in detail by 
\citet{Cernicharo2000}
who derived a $^{12}$C/$^{13}$C value of 45$\pm$3. 
It should be pointed out that for the three
isotopologues, the intensities of the $J$=4-3, and 5-4 transitions differ by less than 2\% and
5\%, respectively  (see Table \ref{tab_hc3n_isomers_irc}). 
Hence, no fractionation effects are observed for HC$_3$N in IRC+10216.

For HC$_5$N, we observed seven lines of each one of its isotopologues towards TMC-1.
Consequently, a rotational diagram can
be built for each one of them (see Table \ref{tab_hc5n_TMC-1}). 
The derived rotational temperatures and column densities are given
in Table \ref{col_den_iso_isomers}. The rotational diagram provides a 
colum density of (9.0$\pm$0.1)$\times$10$^{13}$ cm$^{-2}$, which has
to be considered as a lower limit because the lines show opacity effects.
The weak hyperfine components corresponding to $F=J_u-J_u$ and $J_l-J_l$, which are
placed at $\sim$1.5-2 MHz from the blended three strongest hyperfine components 
$\Delta F$=1, are detected for $J_u$=12 up to 16. As an example, the transition $J=16-15$ has three strong
hyperfine components separated by a few kHz and representing a total line strenght of 47.875.
The two weak hyperfine components have
an added line strenght of 0.125. The theoretical integrated line intensity could have a ratio of 383
between the strong and weak hyperfine components, while the observed value is 153 (see Table \ref{tab_hc5n_TMC-1}). 
For the $J_u$=12 transition, 
the expected ratio is 215, while the observed value is 100.4. Therefore, in order to have a good estimate
of the main isotopologue column density, we have to correct the derived column density from
the strongest hyperfine components by a factor of $\sim$2. Our best estimate for the HC$_5$N column density 
is, therefore, (1.8$\pm$0.2)$\times$10$^{14}$ cm$^{-2}$.

\begin{table*}
\caption{Observed line parameters for HC$_3$N and its isotopologues in TMC-1.}
\label{tab_hc3n_TMC-1}
\centering
\begin{tabular}{{|l|c|c|c|c|c|c|c|c|}}
\hline
Molecule & {\textit J$_u$-J$_l$}& {\textit F$_u$-F$_l$}& $\nu_{rest}^a$& T$_A^*$ & v$_{LSR}^b$   & $\Delta$v$^b$&$\int$T$_A^*$dv $^d$ & $\sigma^e$\\
         &                      &                      &  (MHz)        & (K)     & (kms$^{-1}$)  & (kms$^{-1}$) &  (mK kms$^{-1}$)    &  (mK)\\
\hline         
HCCCN       &4-3& 4-4        & 36390.8876(5)& 0.444&  5.81(01)&0.68(02)&0.3235(10)&0.6\\
HCCCN       &4-3& 3-2        & 36392.2345(5)& 1.865&  5.85(02)&0.65(05)&1.2943(10)&0.6\\
HCCCN       &4-3& 4-3/5-4$^e$& 36392.3429(5)& 2.981&  5.78(02)&0.89(05)&2.8274(10)&0.6\\
HCCCN       &4-3& 3-3        & 36394.1777(5)& 0.457&  5.80(01)&0.67(02)&0.3258(10)&0.6\\
HCCCN       &5-4& 5-5        & 45488.8387(5)& 0.385&  5.81(02)&0.55(03)&0.2358(10)&1.3\\
HCCCN       &5-4& 4-3        & 45490.2594(5)& 1.669&  5.83(10)&0.66(10)&0.8865(10)&1.3\\
HCCCN       &5-4& 5-4/6-5$^e$& 45490.3222(5)& 2.963&  5.81(10)&0.50(10)&2.0868(10)&1.3\\
HCCCN       &5-4& 4-4        & 45492.1107(5)& 0.383&  5.80(02)&0.56(03)&0.2278(10)&1.3\\
H$^{13}$CCCN&4-3& 4-4        & 35265.9791(7)& 0.005&  5.98(05)&0.96(10)&0.0053(10)&0.5\\ 
H$^{13}$CCCN&4-3& 3-2        & 35267.3117(7)& 0.054&  5.80(02)&0.73(02)&0.0415(10)&0.5\\ 
H$^{13}$CCCN&4-3& 4-3/5-4$^e$& 35267.4191(6)& 0.143&  5.76(02)&0.78(02)&0.1185(10)&0.5\\ 
H$^{13}$CCCN&4-3& 3-3        & 35269.2342(7)& 0.006&  5.66(05)&0.68(10)&0.0040(10)&0.5\\ 
H$^{13}$CCCN&5-4$^f$&        & 44084.1622(8)& 0.188&  5.82(02)&0.77(02)&0.1538(10)&1.3\\
HC$^{13}$CCN&4-3& 4-4        & 36236.5139(7)& 0.004&  5.96(05)&0.54(10)&0.0025(10)&0.6\\  
HC$^{13}$CCN&4-3& 3-2        & 36237.8535(7)& 0.064&  5.81(03)&0.65(03)&0.1343(10)&0.6\\  
HC$^{13}$CCN&4-3& 4-3/5-4$^f$& 36237.9619(7)& 0.167&  5.77(03)&0.75(03)&0.1327(10)&0.6\\  
HC$^{13}$CCN&4-3& 3-3        & 36239.7865(7)& 0.006&  5.75(05)&0.63(10)&0.0052(10)&0.6\\  
HC$^{13}$CCN&5-4$^f$&        & 45297.3345(7)& 0.190&  5.83(02)&0.76(02)&0.1532(10)&1.3\\
HCC$^{13}$CN&4-3& 4-4        & 36240.0100(7)& 0.008&  5.85(04)&0.54(02)&0.0048(10)&0.6\\
HCC$^{13}$CN&4-3& 3-2        & 36241.3514(7)& 0.093&  5.82(03)&0.68(02)&0.0674(10)&0.6\\
HCC$^{13}$CN&4-3& 4-3/5-4$^f$& 36241.4436(7)& 0.239&  5.75(03)&0.72(02)&0.1820(10)&0.6\\
HCC$^{13}$CN&4-3& 3-3        & 36243.2862(7)& 0.007&  5.73(04)&0.64(02)&0.0050(10)&0.6\\
HCC$^{13}$CN&5-4$^f$&        & 45301.7069(8)& 0.270&  5.79(01)&0.75(02)&0.2164(10)&1.3\\
HCCC$^{15}$N&4-3 &           & 35333.8879(7)& 0.079&  5.77(02)&0.72(02)&0.0598(10)&0.5\\
HCCC$^{15}$N&5-4 &           & 44167.2678(9)& 0.096&  5.79(04)&0.55(02)&0.0567(10)&1.3\\
DCCCN       &4-3& 4-4        & 33771.0873(3)& 0.007&  5.78(10)&0.78(10)&0.0060(10)&0.5\\
DCCCN       &4-3& 3-2        & 33772.4373(3)& 0.086&  5.74(02)&0.83(03)&0.0754(20)&0.5\\
DCCCN       &4-3& 4-3/5-4$^g$& 33772.5477(3)& 0.229&  5.76(02)&0.79(03)&0.1917(20)&0.5\\  
DCCCN       &4-3& 3-3        & 33774.3772(3)& 0.003&  5.86(10)&0.78(05)&0.0020(05)&0.5\\
DCCCN       &5-4$^f$&        & 42215.5823(9)& 0.320&  5.84(02)&0.78(03)&0.2653(10)&0.9\\
\hline
\end{tabular}
\tablefoot{\\
        \tablefoottext{a}{Adopted rest frequencies (see text).}\\
        \tablefoottext{b}{Local standard of rest velocity of the emission for the adopted rest frequency (in kms$^{-1}$).}\\
        \tablefoottext{c}{Linewidth at half intensity derived by fitting a Gaussian line profile to the observed
     transitions (in kms$^{-1}$).}\\
        \tablefoottext{d}{Integrated line intensity in K kms$^{-1}$. }\\
        \tablefoottext{e}{The sensitivity of the data (root mean square error) has been derived from a baseline to the each line 
      in a velocity window -6.2 to 17.8 km\,s$^{-1}$ (in mK). }\\
    \tablefoottext{f}{Average of the two strongest hyperfine components.}\\
    \tablefoottext{g}{Average of the three strongest hyperfine components.}\\
    
}
\end{table*}

Unlike the case of HC$_3$N, the $^{13}$C isotopologues of HC$_5$N do not show any significant 
abundance ratio anomaly. 
Nevertheless, the isotopologue HCCCC$^{13}$CN shows, for all observed
transitions, a slightly larger intensity than the other $^{13}$C substitutions. In order to evaluate whether this
variation is statistically relevant, we derived the integrated intensity ratios for
each observed transition, taking H$^{13}$CCCCCN as reference. 
The average 
values of these ratios for the $J$=12-11 up to $J$=16-15 lines are:
\\
\\
HC$^{13}$CCCCN/H$^{13}$CCCCCN=0.99$\pm$0.04,
\\
HCC$^{13}$CCCN/H$^{13}$CCCCCN=1.03$\pm$0.04, 
\\
HCCC$^{13}$CCN/H$^{13}$CCCCCN=1.01$\pm$0.04, \\ and
\\
HCCCC$^{13}$CN/H$^{13}$CCCCCN=1.15$\pm$0.04.
\\

\citet{Takano1998} reported the detection of
H$^{13}$CCCCCN and HCCCC$^{13}$CN, claiming an abundance ratio of 1.3 between them. However, our
more sensitive data suggest a similar abundance for all $^{13}$C isotopologues, with a slightly
larger abundance of 15\% (within 3.5$\sigma$) for HCCCC$^{13}$CN. 
The average value for the column density derived from the five $^{13}$C isotopologues 
is (1.94$\pm$0.08)$\times$10$^{12}$ cm$^{-12}$.
Consequently, the $^{12}$C/$^{13}$C abundance ratio in
TMC-1 is 93$\pm$10, which is very close to the solar value. For the observation of two rotational
lines of the five isotopologues, \citet{Taniguchi2016} obtained a similar value for the
isotopic ratio, 94$\pm$6. They also concluded that no significant differences do exist between
the abundances of the five $^{13}$C isotopologues.

Finally, we derived a N/$^{15}$N isotopic abundance ratio of {\textit N}(HC$_5$N)/{\textit N}(HC$_5^{15}$N)=391$\pm$85
and a deuterium enrichment {\textit N}(HC$_5$N)/{\textit N}(DC$_5$N)=82$\pm$20. 
This deuteration enrichment is similar to the one derived from 
HC$_3$N ($\sim$60), HNCCC ($\sim$43), and HCCNC ($\sim$30).

\onecolumn
\begin{longtable}{|l|c|c|c|c|c|c|c|}
\caption[]{Observed line parameters for HC$_5$N, its isotopologues, and HC$_7$N in TMC-1.
\label{tab_hc5n_TMC-1}}\\
\hline 
Molecule & {\textit J$_u$-J$_l$$^a$}& $\nu_{rest}^b$ & T$_A^*$& v$_{LSR}^d$   & $\Delta$v$^e$& $\int$T$_A^*$dv $^c$& $\sigma^f$  \\
         &                          &      (MHz)     & (K)    & (kms$^{-1}$)  & (kms$^{-1}$) &   (K kms$^{-1}$)    &  (mK)  \\
\hline
\endfirsthead
\caption{continued.}\\
\hline
Molecule & {\textit J$_u$-J$_l$$^a$}& $\nu_{rest}^b$ & T$_A^*$& v$_{LSR}^d$   & $\Delta$v$^e$& $\int$T$_A^*$dv $^c$& $\sigma^f$  \\
         &                          &      (MHz)     & (K)    & (kms$^{-1}$)  & (kms$^{-1}$) &   (K kms$^{-1}$)    &  (mK)  \\
\hline
\endhead
\hline
\endfoot
\hline
\endlastfoot
\hline
HC$_5$N        & 12-11& 31951.7764( 4)&1.925&5.83(01)&0.66(01)&1.3574(10)&0.8\\ 
            &$F$=11-11& 31953.4494( 5)&0.008&5.80(04)&0.58(06)&0.0049(10)&0.8\\ 
            &$F$=12-12& 31950.2362( 5)&0.012&5.85(02)&0.69(06)&0.0086(10)&0.8\\ 
HC$_5$N        & 13-12& 34614.3853( 5)&1.667&5.82(01)&0.73(01)&1.2858(10)&0.5\\ 
            &$F$=12-12& 34616.0526( 5)&0.007&5.85(03)&0.63(06)&0.0045(10)&0.5\\ 
            &$F$=13-13& 34612.8403( 5)&0.009&5.77(03)&1.17(07)&0.0105(10)&0.5\\ 
HC$_5$N        & 14-13& 37276.9847( 6)&1.737&5.82(01)&0.66(01)&1.2282(10)&0.5\\ 
            &$F$=13-13& 37278.6473( 5)&0.005&5.78(05)&0.62(11)&0.0031(10)&0.5\\ 
            &$F$=14-14& 37275.4357( 5)&0.006&5.81(04)&0.80(10)&0.0051(10)&0.5\\ 
HC$_5$N        & 15-14& 39939.5741( 6)&1.769&5.84(01)&0.59(01)&1.1078(10)&0.7\\ 
            &$F$=14-14& 39941.2326( 6)&0.005&5.93(05)&0.46(06)&0.0025(10)&0.7\\ 
            &$F$=15-15& 39938.0215( 6)&0.006&5.78(02)&0.76(09)&0.0048(10)&0.7\\ 
HC$_5$N        & 16-15& 42602.1527( 6)&1.710&5.84(01)&0.54(01)&0.9735(10)&1.0\\ 
            &$F$=15-15& 42603.8075( 6)&0.005&5.88(05)&0.49(07)&0.0027(10)&1.0\\ 
            &$F$=16-16& 42600.5968( 6)&0.005&5.92(05)&0.63(12)&0.0037(10)&1.0\\ 
HC$_5$N        & 17-16& 45264.7196( 7)&1.429&5.83(01)&0.55(01)&0.8305(10)&1.3\\
HC$_5$N        & 18-17& 47927.2743( 7)&1.213&5.84(01)&0.57(01)&0.7302(10)&1.6\\ 
\hline
H$^{13}$CCCCCN & 12-11& 31120.0298(22)&0.034&5.85(14)&0.65(03)&0.0236(10)&1.3\\
H$^{13}$CCCCCN & 13-12& 33713.3286(24)&0.031&5.80(04)&0.74(01)&0.0246(10)&0.6\\
H$^{13}$CCCCCN & 14-13& 36306.6185(26)&0.031&5.81(06)&0.64(02)&0.0211(10)&0.6\\
H$^{13}$CCCCCN & 15-14& 38899.8989(27)&0.030&5.82(06)&0.63(01)&0.0201(10)&0.7\\
H$^{13}$CCCCCN & 16-15& 41493.1690(29)&0.028&5.82(08)&0.54(02)&0.0161(10)&0.8\\
H$^{13}$CCCCCN & 17-16& 44086.4282(30)&0.022&5.82(16)&0.56(03)&0.0131(10)&1.2\\
H$^{13}$CCCCCN & 18-17& 46679.6757(32)&0.018&5.79(19)&0.61(04)&0.0117(10)&1.3\\
H$^{13}$CCCCCN & 19-18& 49272.9109(33)&0.018&5.91(29)&0.66(06)&0.0126(10)&1.9\\
\hline
HC$^{13}$CCCCN & 12-11& 31624.3437(20)&0.039&5.88(07)&0.65(01)&0.0270(10)&0.7\\
HC$^{13}$CCCCN & 13-12& 34259.6672(21)&0.032&5.84(05)&0.71(01)&0.0239(10)&0.5\\
HC$^{13}$CCCCN & 14-13& 36894.9814(23)&0.032&5.85(06)&0.59(02)&0.0201(10)&0.6\\
HC$^{13}$CCCCN & 15-14& 39530.2857(24)&0.030&5.88(06)&0.62(02)&0.0192(10)&0.6\\
HC$^{13}$CCCCN & 16-15& 42165.5793(26)&0.026&5.87(10)&0.54(03)&0.0149(10)&0.8\\
HC$^{13}$CCCCN & 17-16& 44800.8615(27)&0.022&5.85(12)&0.48(03)&0.0115(10)&1.2\\
HC$^{13}$CCCCN & 18-17& 47436.1316(28)&0.013&5.86(30)&0.56(07)&0.0078(10)&1.5\\
\hline
HCC$^{13}$CCCN & 12-11& 31918.6866(18)&0.038&5.87(08)&0.65(02)&0.0258(10)&0.7\\
HCC$^{13}$CCCN & 13-12& 34578.5381(19)&0.033&5.86(05)&0.72(01)&0.0256(10)&0.6\\
HCC$^{13}$CCCN & 14-13& 37238.3802(21)&0.033&5.86(07)&0.65(02)&0.0227(10)&0.6\\
HCC$^{13}$CCCN & 15-14& 39898.2122(22)&0.031&5.88(06)&0.60(01)&0.0199(10)&0.7\\
HCC$^{13}$CCCN & 16-15& 42558.0335(23)&0.028&5.87(09)&0.53(02)&0.0156(10)&0.8\\
HCC$^{13}$CCCN & 17-16& 45217.8432(24)&0.025&5.89(15)&0.65(04)&0.0169(10)&1.3\\
HCC$^{13}$CCCN & 18-17& 47877.6406(26)&0.015&5.86(30)&0.49(07)&0.0079(10)&1.7\\
\hline
HCCC$^{13}$CCN & 12-11& 31922.5700(16)&0.036&5.82(08)&0.68(02)&0.0264(10)&0.7\\
HCCC$^{13}$CCN & 13-12& 34582.7452(17)&0.032&5.81(07)&0.69(02)&0.0232(10)&0.6\\
HCCC$^{13}$CCN & 14-13& 37242.9110(18)&0.033&5.83(08)&0.67(02)&0.0234(10)&0.8\\
HCCC$^{13}$CCN & 15-14& 39903.0667(19)&0.030&5.84(06)&0.58(01)&0.0183(10)&0.7\\
HCCC$^{13}$CCN & 16-15& 42563.2117(20)&0.027&5.84(08)&0.56(02)&0.0163(10)&0.9\\ 
HCCC$^{13}$CCN & 17-16& 45223.3451(21)&0.023&5.81(12)&0.53(03)&0.0128(10)&1.1\\
HCCC$^{13}$CCN & 18-17& 47883.4663(22)&0.021&5.86(27)&0.61(06)&0.0135(10)&1.7\\
\hline
HCCCC$^{13}$CN & 12-11& 31636.1315(14)&0.042&5.86(07)&0.66(02)&0.0294(10)&0.8\\
HCCCC$^{13}$CN & 13-12& 34272.4373(15)&0.036&5.81(05)&0.72(01)&0.0277(10)&0.5\\
HCCCC$^{13}$CN & 14-13& 36908.7338(16)&0.038&5.84(05)&0.66(01)&0.0263(10)&0.6\\
HCCCC$^{13}$CN & 15-14& 39545.0204(17)&0.033&5.85(05)&0.61(01)&0.0216(10)&0.7\\
HCCCC$^{13}$CN & 16-15& 42181.2963(18)&0.034&5.85(06)&0.54(02)&0.0193(10)&0.8\\
HCCCC$^{13}$CN & 17-16& 44817.5608(19)&0.025&5.84(11)&0.60(03)&0.0156(10)&1.0\\
HCCCC$^{13}$CN & 18-17& 47453.8132(20)&0.020&5.86(16)&0.63(04)&0.0132(10)&1.3\\
\hline
HCCCCC$^{15}$N & 12-11& 31167.1619(15)&0.010&5.90(35)&0.61(06)&0.0064(10)&0.9\\
HCCCCC$^{15}$N & 13-12& 33764.3882(17)&0.007&5.80(24)&0.64(07)&0.0045(10)&0.5\\
HCCCCC$^{15}$N & 14-13& 36361.6056(18)&0.010&5.75(21)&0.81(05)&0.0084(10)&0.6\\
HCCCCC$^{15}$N & 15-14& 38958.8134(19)&0.010&5.75(22)&0.83(05)&0.0087(10)&0.6\\
HCCCCC$^{15}$N & 16-15& 41556.0109(20)&0.007&5.81(33)&0.57(07)&0.0042(10)&0.8\\
HCCCCC$^{15}$N & 17-16& 44153.1975(21)&0.007&5.84(44)&0.53(10)&0.0039(10)&1.2\\
HCCCCC$^{15}$N & 18-17& 46750.3723(22)&0.007&5.79(27)&0.54(08)&0.0039(10)&1.2\\
\hline
DC$_5$N        & 13-12& 33047.2676(05)&0.040&5.84(01)&0.66(02)&0.0284(10)&0.6\\
DC$_5$N        & 14-13& 35589.3244(05)&0.037&5.86(01)&0.66(01)&0.0260(10)&0.6\\
DC$_5$N        & 15-14& 38131.3723(05)&0.036&5.86(01)&0.62(02)&0.0238(10)&0.5\\
DC$_5$N        & 16-15& 40673.4104(06)&0.038&5.87(01)&0.51(01)&0.0205(10)&0.7\\
DC$_5$N        & 17-16& 43215.4382(06)&0.029&5.89(01)&0.49(02)&0.0151(10)&1.0\\
DC$_5$N        & 18-17& 45757.4551(07)&0.022&5.88(02)&0.70(05)&0.0167(10)&1.3\\
DC$_5$N        & 19-18& 48299.4603(07)&0.016&5.79(04)&0.56(07)&0.0097(10)&1.7\\
\hline
\hline
HC$_7$N        &28-27 & 31583.7088(10)&0.320&5.85(02)&0.63(01)&0.2147(10)&0.8\\ %
HC$_7$N        &29-28 & 32711.6717(10)&0.272&5.84(02)&0.67(01)&0.1946(10)&0.6\\ %
HC$_7$N        &30-29 & 33839.6318(10)&0.229&5.83(02)&0.71(01)&0.1722(10)&0.5\\ %
HC$_7$N        &31-30 & 34967.5890(10)&0.203&5.85(02)&0.69(01)&0.1480(10)&0.5\\ %
HC$_7$N        &32-31 & 36095.5431(11)&0.189&5.86(02)&0.64(01)&0.1298(10)&0.7\\ %
HC$_7$N        &33-32 & 37223.4942(11)&0.168&5.85(02)&0.63(01)&0.1134(10)&0.6\\ %
HC$_7$N        &34-33 & 38351.4420(12)&0.147&5.86(02)&0.60(02)&0.0938(10)&0.7\\ %
HC$_7$N        &35-34 & 39479.3866(12)&0.124&5.87(02)&0.58(02)&0.0766(10)&0.6\\ %
HC$_7$N        &36-35 & 40607.3277(12)&0.113&5.87(02)&0.52(02)&0.0624(10)&0.9\\ %
HC$_7$N        &37-36 & 41735.2654(12)&0.092&5.87(03)&0.49(02)&0.0476(10)&1.1\\ %
HC$_7$N        &38-37 & 42863.1995(13)&0.077&5.87(03)&0.51(03)&0.0415(10)&0.9\\ %
HC$_7$N    &39-38$^g$ & 43991.1299(13)&     &        &        &          &1.0\\ %
HC$_7$N        &40-39 & 45119.0565(13)&0.047&5.87(02)&0.54(03)&0.0270(10)&1.4\\ %
HC$_7$N        &41-40 & 46246.9792(13)&0.043&5.87(02)&0.46(03)&0.0210(10)&1.3\\ %
HC$_7$N        &42-41 & 47374.8980(14)&0.022&5.88(03)&0.49(04)&0.0116(10)&1.5\\ %
HC$_7$N        &43-42 & 48502.8127(14)&0.018&5.78(02)&0.64(06)&0.0125(10)&1.7\\ %
HC$_7$N        &44-43 & 49630.7232(14)&0.013&5.95(06)&0.57(12)&0.0076(10)&2.1\\ %
\hline
\end{longtable}
\tablefoot{\\
    \tablefoottext{a}{Except when indicated, it corresponds to the group of
     the three strongest hyperfine components ($\Delta$F=1).}\\ 
        \tablefoottext{b}{Adopted rest frequencies (see text).}\\
        \tablefoottext{c}{Local standard of rest velocity of the emission for the adopted 
    rest frequency (in kms$^{-1}$).}\\
        \tablefoottext{d}{Linewidth at half intensity derived by fitting a Gaussian line profile to the observed
     transitions (in kms$^{-1}$).}\\
        \tablefoottext{e}{Integrated line intensity in K kms$^{-1}$.}\\
        \tablefoottext{f}{The sensitivity of the data (root mean square error) has been derived 
     from a baseline to the each line in a velocity window -6.2 to 17.8 km\,s$^{-1}$ (in mK). }\\
        \tablefoottext{g}{Line fully blended with a strong negative feature produced by the
     frequency switching observing procedure.}\\
}

\twocolumn
\section{Quantum chemical calculations for the  HC$_5$N isomers}
\label{App_ab_initio}

\begin{figure}[]
        \label{ener_isomers}
        \centering
        \includegraphics[scale=0.3,angle=0]{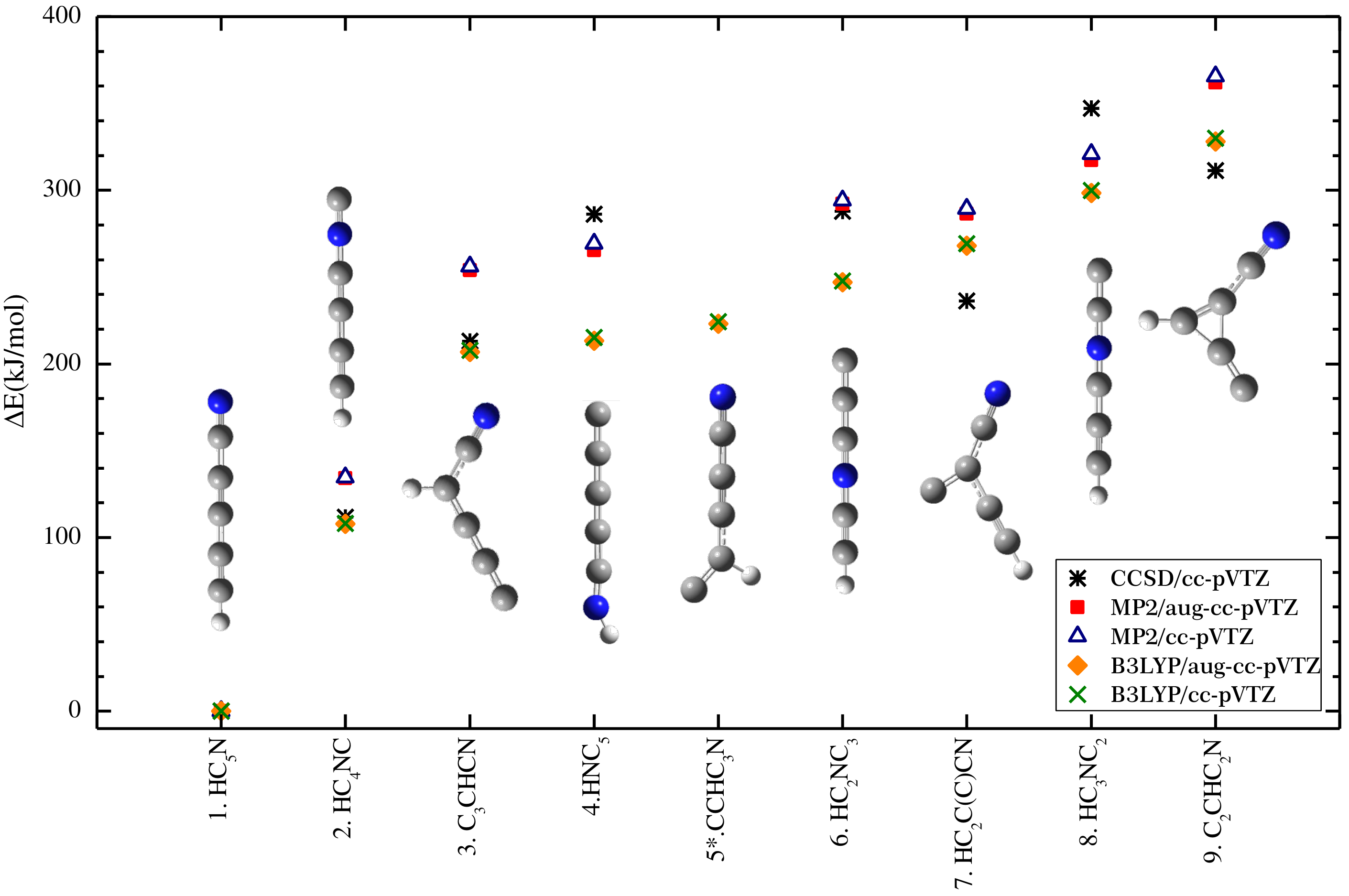}
        \caption{Theoretical energies found at different levels of theory for the 
    HC$_{5}$N isomers that stand below 500~kJ/mol. The isomers are represented from the 
    optimized coordinates obtained at MP2/cc-pVTZ level of theory, except for those marked 
    with an asterisk in their name, where the B3LYP/aug-cc-pVTZ structures were used, see 
    text of Appendix \ref{App_ab_initio}. }
\end{figure}

The relative stability between the possible isomers of HC$_{5}$N has been evaluated 
by different quantum chemical calculations. As initial values for our calculations, 
we took the optimized structures present in the literature (\citealp{Gronowski2006,Gronowski2007}). 
We optimized these structures using sequential calculation methods in oder of complexity: first, 
the density functional theory (DFT) hybrid exchange-correlation variant (B3LYP), then the M{\o}ller-Plesset 
 second-order perturbation theory \citep{Moller1934}, and, finally, closed-shell coupled cluster with 
singles and doubles (CCSD) \citep{Cizek1969}. For all these methods, we used the Dunning's 
correlation consistent polarized valence triple-$\zeta$ basis sets (cc-pVTZ), and, in the case of the 
DFT and MP2 methods, we increased the basis sets with diffuse functions (aug-cc-pVTZ), see 
Figure \ref{ener_isomers}.1. Besides the optimization, we have calculated some relevant parameters 
such as the vibrational energies, IR intensities and the rotational constants at the vibrational 
ground state, thanks to the frequency calculations using the MP2/cc-pVTZ level of theory under 
anharmonic corrections (see Table \ref{table_calculations}). All these calculations 
were performed using the Gaussian 09 \citep{Frisch2013} software package.

The energy order of the HC$_{5}$N isomers does not drastically change as a function of the 
level of theory assumed. As expected, the two most stable isomers are HC$_{5}$N (\underline{\textbf{1}}) and 
HC$_{4}$NC (\underline{\textbf{2}}), separated in energy by $\sim$110~kJ/mol. In the 200 -400~kJ/mol range, 
we found seven different species, including: additional linear structures (or quasi linear) such as:  HNC$_{5}$ 
(\underline{\textbf{4}}),  HC$_{2}$NC$_{3}$ (\underline{\textbf{6}}), and HC$_{3}$NC$_{2}$ (\underline{\textbf{8}});
 a cyclic species (C$_{2}$CHC$_{2}$N (\underline{\textbf{9}}); and 
branched structures: C$_{3}$CHCN (\underline{\textbf{3}}), CCHC$_{3}$N (\underline{\textbf{5}})), 
HC$_{2}$C(C)CN (\underline{\textbf{6}}). In case of the quasi linear structure HNC$_{5}$ (\underline{\textbf{4}}), 
we  verified the non-linearity of its structure by calculating the potential well; see Figure \ref{C5NHlinear}.2. 
As pointed out in the literature \citep{Gronowski2006}, some branched isomers are more stable that the linear 
structures. Nevertheless, we found some difficulties in finding the minimum energy potential for some of these
branched species: 
\underline{\textbf{5}} and \underline{\textbf{7}}.

\begin{figure}[]
        \label{C5NHlinear}
        \centering
        \includegraphics[scale=0.305]{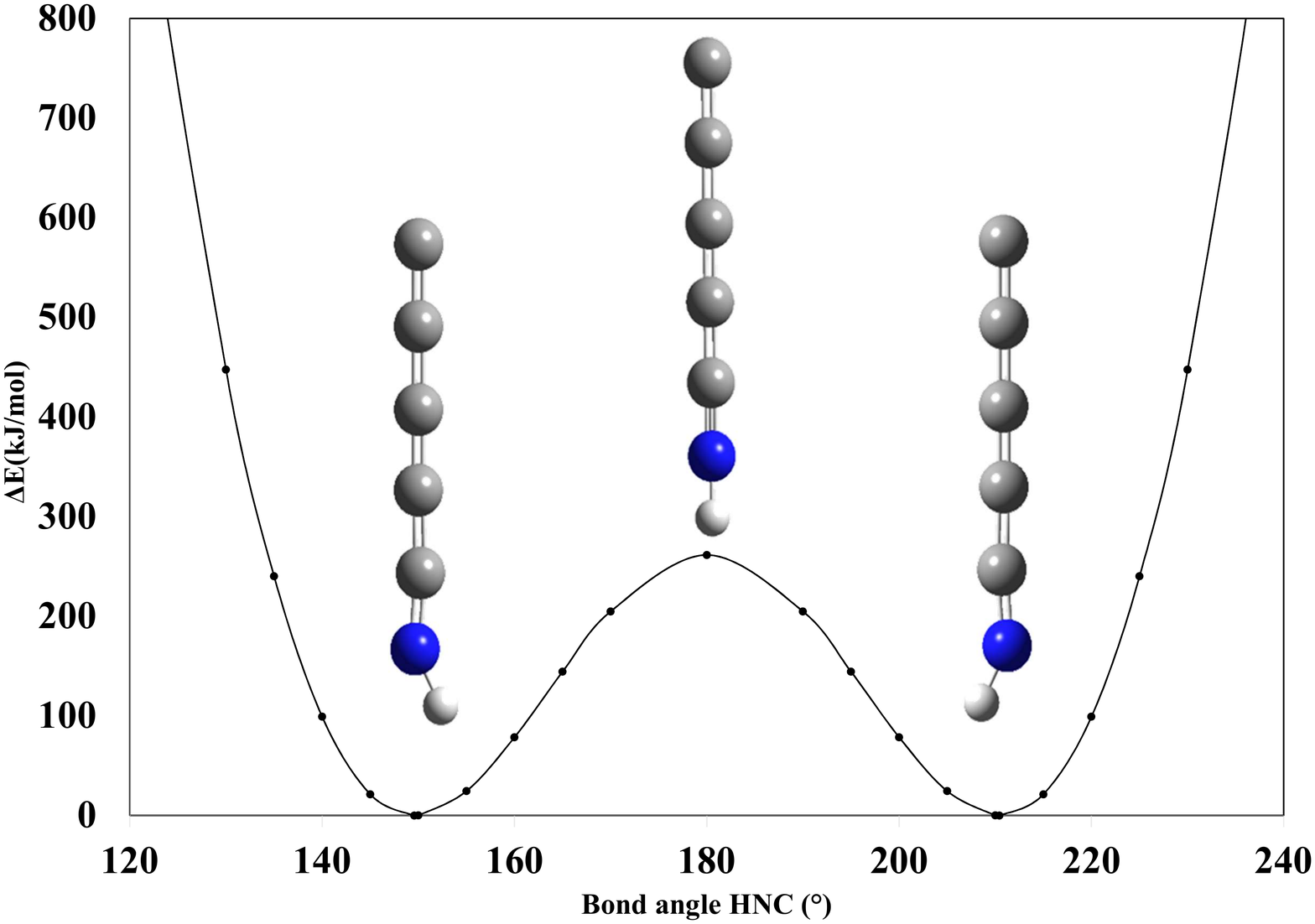}
        \caption{Potential energy well obtained by evaluating the $HNC$ bond angle of the 
    HNC$_{5}$ species (\underline{\textbf{4}}).}
        
\end{figure}

Species \underline{\textbf{5}},  CCHC$_{3}$N, was considered the one with lowest energy among the 
branched C$_{5}$HN isomers in the literature (\citealp{Gronowski2006,Gronowski2007}). The relative stability of  
CCHC$_{3}$N compared to other linear structures has led a number of authors to propose it, together with 
the two carbon less analogue, CCHCN, as good candidates for searches in space. In our calculations, 
we could find the optimized structure \underline{\textbf{5}} using the B3LYP method, however, neither 
the MP2 or CCSD converged to this structure. To understand the non-convergence to this structure, we 
analyzed the potential energy along the C$_3$ C$_4$ C$_5$ bond angle and we compared the 
results between the B3LYP and MP2 methods; see Figure \ref{CCHC3N_scan}.3. For the structure in the 
B3LYP calculation, we found a minimum at a C$_3$ C$_4$ C$_5$ bond angle of 137$^{\circ}$. In 
contrast, for the MP2 calculations, any potential well was found around this angle. 
The only stable structure obtained with MP2 and CCSD methods corresponds to a species where
the hydrogen is no longer 
covalently linked to the carbon chain which is linearly bounded. 

\begin{figure}[]
        \label{CCHC3N_scan}
        \centering
        \includegraphics[scale=0.305]{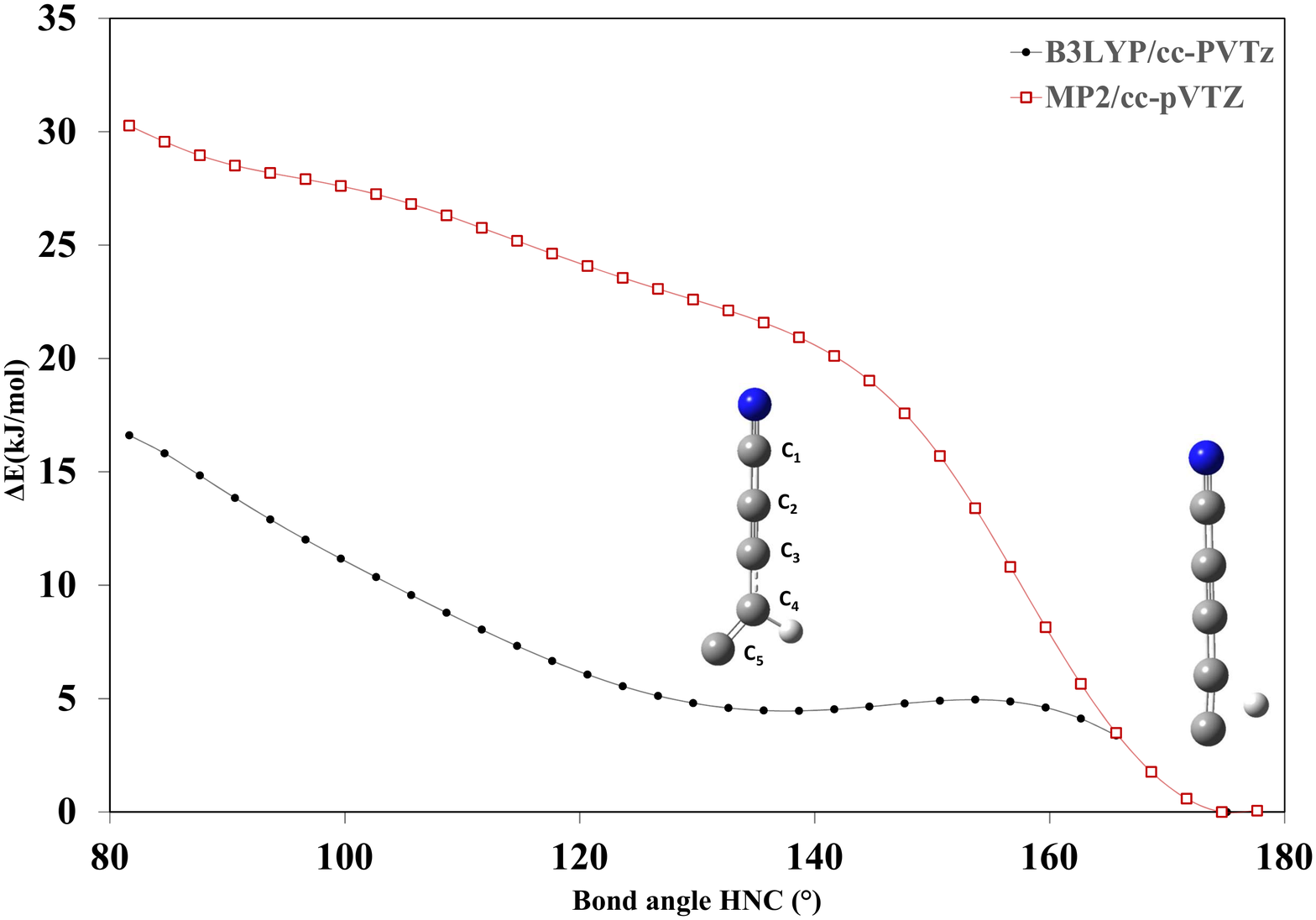}
        \caption{Potential energy scan of the C$_3$ C$_4$ C$_5$ bond
          angle for the \underline{\textbf{5}} species (CCHC$_{3}$N) evaluated at MP2/cc-pVTZ
          and B3LYP/cc-pVTZ levels of theory.}
\end{figure}
     
Another isomer that should be pointed out is the species \underline{\textbf{7}},HC$_{2}$C(C)CN
since the
bond angle around the branching structure C$_4$ C$_2$ C$_3$ has a non-conventional value of 
$\sim$90$^{\circ}$.
A similar structure was proposed between the most stables isomers by \cite{Gronowski2006}.
However, calculations by the same author in 2007 indicated that
the bond angle of the optimized structure was $\sim$120$^{\circ}$ \citep{Gronowski2007}. 
Our
scan along the C$_4$ C$_2$ C$_3$ bond angle (Figure \ref{HC2C-C-CN_scan}.4) at MP2/cc-pVTZ level of
theory does not show any potential minima at the $\sim$120$^{\circ}$ angle. The only structures that
converged in the calculations are those corresponding to zero degrees, that is, the
species \underline{\textbf{1}}, and the $\sim$90$^{\circ}$ one. Neither 
the cyclic structure or the $\sim$120$^{\circ}$ converged with the MP2 methodology.
In contrast, CCSD calculations show that the real minima is at $\sim$120$^{\circ}$ bond angle, which
is the reason why the energy at CCSD is lower than the one at MP2 calculations.

Independently of the level of theory employed in the calculations, the two observed species 
in space, HC$_{5}$N (\underline{\textbf{1}}) and HC$_{4}$NC (\underline{\textbf{2}}), are the 
most stable isomers of cyanodiacetylene.

 The rotational constants we have calculated for the different isomers of HC$_5$N are 
 precise enough to search for them in space. However, we could expect very weak lines, which would be well 
  below those of HC$_4$NC. Hence, the possibility to detect them in interstellar clouds 
  without precise laboratory frequencies is rather unlikely.
     
\begin{figure}[]
        \label{HC2C-C-CN_scan}
        \centering
        \includegraphics[scale=0.305]{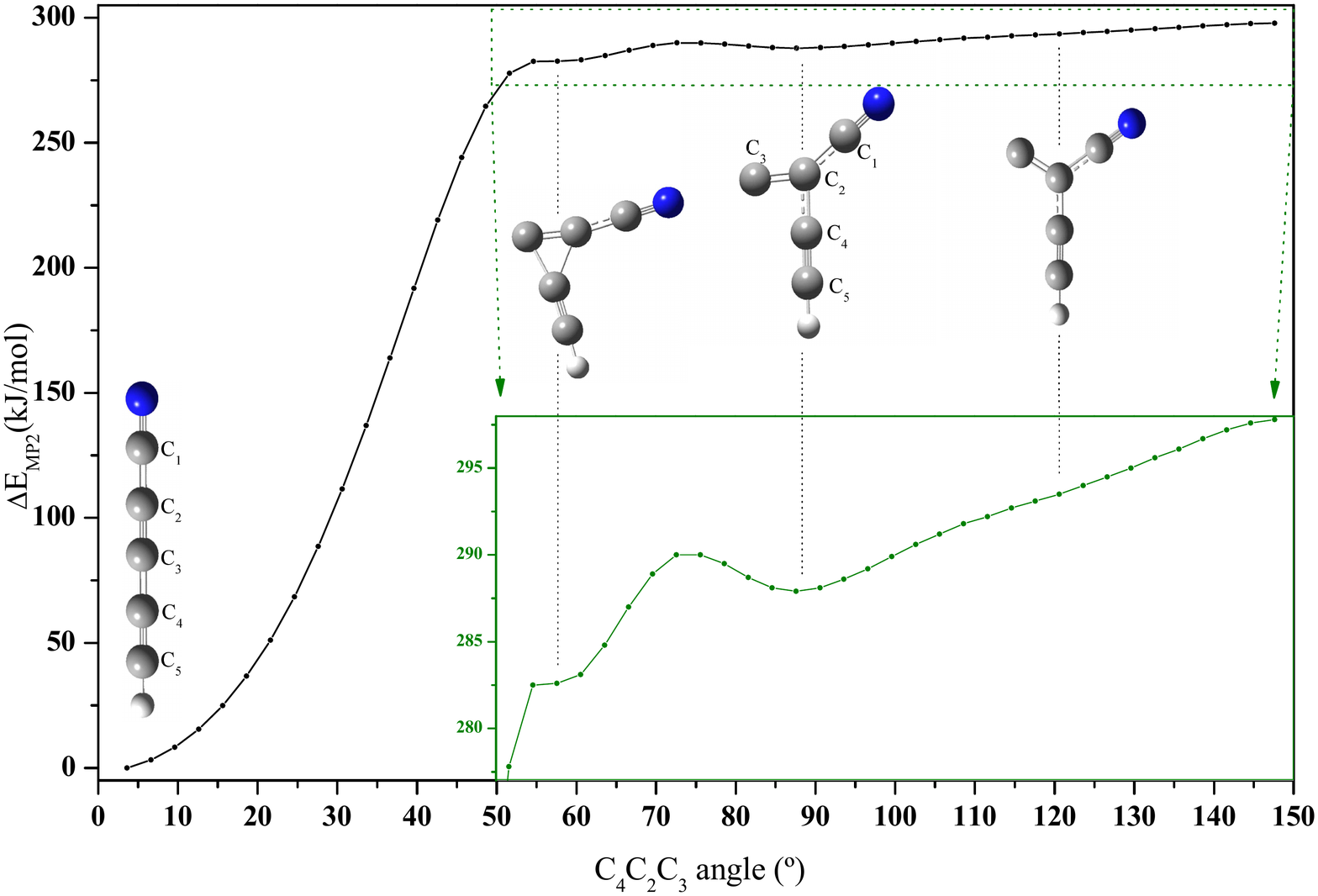}
        \caption{Potential energy scan of the C$_4$ C$_2$ C$_3$ bond angle for the isomer 
    HC$_{2}$C(C)CN (\underline{\textbf{6}}) evaluated at MP2/cc-pVTZ. }
        
\end{figure}

\begin{table*}
        \caption{Energies, rotational constants, quadrupole coupling constants, dipole moment, and vibrational partition functions of the HC$_{5}$N isomers evaluated at MP2/cc-pVTZ level of theory.}
        \label{table_calculations} 
        \begin{tabular}{{lcccccccccc}}
                \hline\hline   
                & $\Delta E$\tablefootmark{(a)}  & $\Delta E_{MP2}$\tablefootmark{(b)} & $A_{0}$\tablefootmark{(c)} & $B_{0}$\tablefootmark{(c)} & $C_{0}$\tablefootmark{(c)} & $1.5\chi_{aa}$
                & $0.25(\chi_{bb}-\chi_{cc})$ & $\mu_{T}$ & $Q_{v}$\tablefootmark{(d)} & $Q_{v}$\tablefootmark{(d)} \\ 
                & (kJ/mol) & (kJ/mol) & (MHz) & (MHz) & (MHz) & (MHz) & (MHz) & (D) & 50K  & 300K  \\ \hline
                HC$_{5}$N$^e$     & 0.0   & 0.0   &              & 1320.046  &           & -5.77 &       & 4.36 & 1.09 & 22.89 \\
                HC$_{4}$NC$^f$    & 134.7 & 133.8 &              & 1390.737  &           & ~1.15 &       & 3.57 & 1.07 & 22.62 \\
                C$_{3}$CHCN       & 256.1 & 254.5 & 21076.776    & 1689.938  & 1564.496  & -2.73 & -0.41 & 3.09 & 1.10 & 14.18 \\
                C$_{5}$NH         & 269.2 & 267.7 &              & 1348.267  &           & ~2.10 & ~0.31 & 9.14 & 1.02 & ~8.86 \\
                HC$_{2}$NC$_{3}$  & 294.1 & 292.2 &              & 1434.147  &           & ~2.96 &       & 6.58 & 1.01 & ~8.06 \\
                HC$_{2}$C(C)CN    & 289.3 & 284.2 & 10693.657    & 2526.651  & 2038.799  & -4.72 & -0.20 & 3.27 & 1.06 & 15.93 \\
                HC$_{3}$NC$_{2}$  & 320.8 & 316.2 &              & 1435.856  &           & ~0.91 &       & 7.15 & 1.03 & 15.46 \\
                C$_{2}$CHC$_{2}$N & 365.5 & 361.3 & 11195.360    & 2552.013  & 2073.245  & -5.24 & -0.14 & 4.48 & 1.15 & 15.16 \\
                \hline                                                                                                                                                                                                                                                                                                                                                                                                                                               
        \end{tabular}   
        \tablefoot{                                                                   
                \tablefoottext{a}{Relative energy to the lowest energy isomer HC$_{5}$N evaluated at MP2/cc-pVTZ level of theory.}\\
                \tablefoottext{b}{Relative energy including the zero-point correction energy calculated for each species.} \\
                \tablefoottext{c}{Rotational constants at the ground state evaluated using the vibration-rotation interaction constants obtained using the anharmonic corrections.
                        In the case of the quasi linear C$_{5}$NH, the effective rotational constant, $B_{eff}$, is provided.}\\
                \tablefoottext{d}{Vibrational partition function calculated using the expression    
        $\prod_{i}(1-e^{-(h\omega_{i}/KT)})^{-d_{i}}$ \citep{Gordy}, where $\omega_{i}$ and $d_{i}$ represent the energy 
        and the degeneracy of each $i$ vibrational mode. The energies for the vibrational modes are taken from the results 
        of ab initio calculations evaluated at MP2/cc-pVTZ level of theory, under the anharmonic correction. 
        At the temperature of TMC1 (~10K), the vibrational partition function for isomers is equal to one.}\\
        \tablefoottext{e}{The experimental value is $B_0$=1331.33269$\pm$0.00002 \citep{Bizzocchi2004}. The
        ratio $B_0(exp)/B_0(Cal)$ is 1.0086.}\\
        \tablefoottext{f}{The experimental value is $B_0$=1401.18216$\pm$0.00006 (this work, see Table \ref{new_rot_const}.
        The ratio $B_0(exp)/B_0(Cal)$ is 1.0075.}\\
        
        }
\end{table*}

\end{appendix}

\end{document}